\documentclass{conm-p-l}
\usepackage{algorithm}
\usepackage[noend]{algpseudocode}
\usepackage{tikz-cd}
\usepackage{lineno}
\usepackage{amsmath}
\usepackage{amsthm}
\linenumbers
\usepackage[symbol]{footmisc}
\usepackage{graphicx}
\usepackage{pst-node}
\usepackage{tikz-cd} 
\usepackage{mathtools}
\usepackage{amssymb}
\usepackage{amsmath}
\usepackage{caption}
\usepackage{amsthm}
\usepackage{array}
\usepackage{float}
\usepackage{mathtools}
\usepackage{graphicx}
\usepackage{enumitem}
\usepackage{hyperref}
\usepackage[numbers]{natbib}   
\usepackage{epigraph}
\usepackage{MnSymbol}

\usepackage{dsfont}
\usepackage{graphicx}
\usepackage{caption}
\usepackage{subcaption}
\usepackage{relsize}
\usepackage{todonotes}
\usepackage{lineno}
\nolinenumbers
\newtheorem{theorem}{Theorem}[section]
\newtheorem{lemma}[theorem]{Lemma}

\theoremstyle{definition}
\newtheorem{definition}[theorem]{Definition}

\theoremstyle{remark}

\numberwithin{equation}{section}

\begin{document}

\title[CY metrics via Grassmannian learning and Donaldson's algorithm]{Calabi--Yau metrics through Grassmannian learning and Donaldson's algorithm}


%    author one information
\author{Carl Henrik Ek}
\address{The Computer Laboratory, University of Cambridge, Cambridge, UK}
\address{Karolinska Institutet, Stockholm, Sweden}
\curraddr{}
\email{che29@cam.ac.uk}
\thanks{Published in \emph{Contemporary Mathematics} \textbf{835}
(2026), American Mathematical Society.
\url{https://doi.org/10.1090/conm/835}.}

%    author two information
\author[Oisin Kim]{Oisin Kim}
\address{The Computer Laboratory, University of Cambridge, Cambridge, UK}
\curraddr{}
\email{osmk3@cam.ac.uk}

\author{Challenger Mishra}
\address{The Computer Laboratory, University of Cambridge, Cambridge, UK}
\curraddr{}
\email{cm2099@cam.ac.uk}
\thanks{}

%    \subjclass is required.
\subjclass[2020]{Primary 14J32}

\date{}

%    Abstract is required.
\begin{abstract}
Motivated by recent progress in the problem of numerical K\"ahler metrics, we survey machine learning techniques in this area, discussing advantages and drawbacks. We then revisit the algebraic ansatz pioneered by Donaldson. Inspired by his work, we present a novel approach to obtaining Ricci-flat approximations to K\"ahler metrics, applying machine learning within a principled framework that guarantees positivity of the metric and thus K\"ahlericity. This offers a solution to a problem arising from the use of machine-learned metrics. In particular, we use gradient descent on the Grassmannian manifold to identify an efficient subspace of sections for calculation of the metric. We combine this approach with both Donaldson's algorithm and learning on the $h$-matrix itself (the latter method being equivalent to gradient descent on the fibre bundle of Hermitian metrics on the tautological bundle over the Grassmannian). We implement our methods on the Dwork family of threefolds, commenting on the behaviour at different points in moduli space. In particular, we observe the emergence of nontrivial local minima as the moduli parameter is increased.
\end{abstract}

\maketitle

\section{Introduction}

At the $1954$ International Congress of Mathematicians, the geometer Eugenio Calabi proposed a conjecture that would dominate his field for decades. In 1933 Erich K\"ahler had defined K\"ahler manifolds, combining a compatible Riemannian, complex and symplectic structure \cite{Kahler}. Twenty years later, Calabi argued that the Ricci curvature of such a manifold should be arbitrarily prescribable, assuming the most naive topological restriction \cite{Calabi}. More specifically, given a symmetric $(0,2)$ tensor $h$, one can always ask whether there exists a metric $g$ on a Riemannian $M$, such that $Ric_g = h$. In general this is a highly difficult problem, involving a collection of nonlinear PDEs. However, Calabi argued that on K\"ahler manifolds, it is sufficient to consider the topological first Chern class alone. Since this is represented by the normalised Ricci form associated to the metric, we obtain a necessary condition on $h$, that the appropriately normalised $(1,1)$-form associated to it should also represent the first Chern class. Calabi's remarkable claim was that this obviously necessary condition is also sufficient. 

Whilst this may seem to be an esoteric mathematical problem, it becomes easily motivated from a physical perspective by considering the vacuum Einstein field equations (EFE). These are expressible in the form $\rho=0$ on a K\"ahler manifold with vanishing Chern class, so that a solution to the prescribed Ricci curvature problem with $h=0$ is also one for the vacuum EFE. Underlying this discussion are remarkable aspects of K\"ahler geometry, most importantly that the Ricci curvature is locally expressible in terms of a real-valued function. Calabi's conjecture was believed false by leading geometers for several decades, as described by S.T. Yau, the mathematician who eventually proved it \cite{hosono}, \cite{yau_2001}. This was because a merely topological condition was believed to be insufficient for the existence of the desired metrics. Yau began his work by seeking a counterexample, and has the following remarkable quote on the time he spent searching for one: ‘Every time I gave one, it failed in a very delicate manner, so I felt it cannot be that delicate unless God had fooled me; so it had to be right... I changed my mind completely, and then I prepared everything to try to solve it’ \cite{hosono}. He succeeded in 1976 \cite{Yau}, and was awarded the Fields medal in 1982, largely in recognition of this achievement. 

Once the existence and (appropriate) uniqueness of Ricci-flat metrics had been established in the case of vanishing Chern class, such manifolds became known as Calabi-Yaus. This fit into a wider programme of modernising and classifying Riemannian geometries, for example by holonomy groups. For Riemannian manifolds these are contained in the orthogonal group, since the metric $g_{ij}$ is parallel with respect to the Levi-Civita connection. At the same 1954 conference (also attended by Cartan and Dolbeault, amongst others), Marcel Berger had presented his famous classification \cite{BSMF_1955__83__279_0}. In the K\"ahler case the holonomy group is contained inside the unitary group, because the K\"ahler form is parallel. For Calabi-Yaus, this is restricted to the special unitaries, essentially because the nowhere-vanishing $(n,0)$-form $\Omega$ associated to $M$ is also parallel. 

During the mid 1980s the field was again revolutionised by theoretical physics, in particular by the landmark formulation of $6+4=10$-dimensional superstring theory \cite{Candelas}. In its framework, spacetime is modelled as a manifold of $10$ real dimensions, with the $6$ extra dimensions forming a compact manifold $M$ which is in some sense small \cite{Douglas_2015}. It turns out that $M$ must itself be Einstein, with the assumption of zero vacuum energy. Since it must also be without infinitesimal isometries\footnote{This would lead to the observation of particles that do not occur in nature.}, the only known choice of compactifying spaces are three-dimensional compact complex K\"ahler manifolds which are Ricci-flat, i.e. Calabi-Yau threefolds. An `industry' for the production of Calabi-Yaus was developed, one of its goals being the identification of properties necessary for recovering observed physics. For example, to obtain the the correct gauge group for the standard model, $SU(3) \times SU(2) \times U(1)$, one requires more conditions on $M$, including the existence of a holomorphic vector bundle $\mathcal{V}$ with a Hermitian Yang-Mills connection. 

Given a Calabi-Yau (CY) $M$, the goal is to calculate meaningful physical quantities, such as accessible particle masses and Yukawa couplings associated to the low energy theory, from the compactified heterotic string theory. The relevant calculations are integrals on $M$; the corresponding integrands are the wedge products of bundle-valued harmonic forms \cite{Douglas_2015}. In the case that the bundle is the tangent bundle, that is, $\mathcal{V}=TM$, known as the standard embedding, the Yukawa couplings are integrals of wedged $(1,1)$ and $(2,1)$-forms. There are special situations in which physical data may be derived topologically \cite{PhysRevLett.55.2547}, since the integrals depend only on cohomology classes of the modes. However, the physically meaningful normalised coupling constants require a correct choice of representative, necessitating knowledge of the Ricci-flat metric. Yau's existence proof was not constructive. This means that outside cases where alternative techniques (e.g. special geometry) can be applied, many physically interesting cases are likely to require numerical approximation.

Beyond phenomenology, the discovery and classification of Calabi-Yaus led to significant cross-fertilisation between mathematics and high-energy physics, centred on the concept of mirror symmetry. In its
simplest form, this conjectures that for every Calabi-Yau threefold $M$, there exists a mirror manifold $\tilde{M}$, such that the Hodge diamonds, diagrams describing the structures of the Dolbeault cohomologies, are related by a mirror reflection along the diagonal \cite{Hori:2003ic}. On a Calabi-Yau threefold with $h^{1,0}=0$, the only nontrivial parts of the Hodge diamond are $h^{1,1}$ and $h^{2,1}$, interchanged by this map. Since these cohomologies correspond to deformations of the symplectic and complex structures, mirror symmetry can also be formulated as a duality between this pair. Whilst further discussion of this profound area is beyond the scope of this paper, it displays the rich interplay between theoretical physics and pure mathematics in the second half of the twentieth century. A major player in this was Simon Donaldson, the next foundational figure of this paper.

Whilst rapid progress was made on the theoretical side, the numerical computation of Ricci-flat metrics lagged behind. The first results were in the work of Headrick and Wiseman \cite{K3}, who approximated the K\"ahler potential, a local real-valued function encoding the same information as the metric. However, computations took on the order of several days due to memory requirements. A significant breakthrough was the work of Donaldson \cite{Donaldson}, \cite{Scalar}, \cite{Scalarii}, building on ideas outlined by Yau, Tian, and others \cite{tian}, \cite{Open}, \cite{zelditch}, \cite{Catlin1999}, \cite{lu}. Donaldson had proven a sequence of significant results in 1980s by using ideas from physics, in particular Yang Mills gauge theory, to study the topology of four manifolds. In the case of constant scalar curvature K\"ahler (cscK) metrics, previous work had introduced approximation schemes via sequences of embeddings into projective space, using the sections of a holomorphic line bundle $\mathcal{L}$. Donaldson's first contribution was to prove a convergence guarantee in the general cscK case, assuming that the automorphism group $Aut(M, \mathcal{L})$ is discrete. Secondly, specialising to the case of Calabi-Yaus, he provided a simplified method, based on the existence of the CY volume form. His elegant work was developed in \cite{Braun_2008} by Douglas and collaborators, who implemented the theory on a specific class of quintics.

Whilst Donaldson's method represented a triumph of inventive computational algebraic geometry, in practice it was limited by the `curse of dimensionality', associated to the growth of the space of global holomorphic sections. Another method, using energy functional minimisation within his framework, was soon introduced \cite{headrick}. Of course, all such research was being conducted in a time of comparatively limited computational resources. The explosive growth of hardware and computable data in recent decades has enabled rapid advances in data science and machine learning (ML). From the start, these have had a major influence in the applied natural sciences \cite{hogg2024machinelearninggoodbad}. On the other hand, applications in pure mathematics and theoretical physics have not been as significant, although this is beginning to shift. Generalising greatly, there are fairly obvious reasons for this: it is hard to apply black-box models with the potential for false positives in disciplines focused on rigour and explicability. 

The computation of Ricci-flat metrics lies on this spectrum. Despite the sophisticated mathematical setup, ultimately it involves approximating a matrix-valued function on a manifold. This is evaluated with a well-defined loss, the integral of scalar-valued quantities associated to the curvature. However, care must be taken when working with K\"ahler metrics, requiring consideration of the individual mathematical constraints, as well as unique difficulties arising from their simultaneous enforcement. Inspired by other ML approaches to PDEs, such as physics-informed neural networks (PINNs) \cite{RAISSI2019686}, a series of recent collaborations have applied ML techniques to the CY metrics problem \cite{moduli}, \cite{ashmore2020machine}, \cite{challenger}, \cite{Larfors:2022nep}, 
\cite{pmlr-v145-douglas22a}, \cite{berglund2023machine}, \cite{mirjanić2025symbolicapproximationsricciflatmetrics}, \cite{Hendi:2024yin}. The basic idea is to impose the constraints on the metric and its derivatives in the most straightforward way, composing them with the Ricci-flatness objective into a single loss function, alongside designing neural architectures that automatically incorporate some of the constraints. This work has culminated in the computation of exact Yukawa couplings with respect to both standard and non-standard embeddings \cite{butbaia2024physicalyukawacouplingsheterotic},  \cite{Constantin:2024yxh}. 
In the first case, significant agreement is found with results obtained by alternative methods. The importance of such results should not be downplayed; as stated in a recent article summarising progress in this research direction, `until now, any such calculations would have been unthinkable' \cite{quanta}.

Contemporary ML work has obvious advantages over past approaches, leveraging the usual techniques, such as parallel computing and efficient optimisation via backpropagation, to compute metrics with high degrees of accuracy with relatively low computational cost. However, in this paper, we will argue that despite very real progress, there are currently unaddressed drawbacks associated to error composition. Most importantly, there are serious questions surrounding the positivity of the metrics, fundamental to their very definition. Whilst the current literature has moved far beyond basic computation into actual physical geometry applications, we suggest here that a rush to calculate must be accompanied by a consistent reexamination of the fundamentals. 

The paper is structured as follows. After a review of the necessary mathematical framework, we discuss ML approaches to CY metrics in some detail, presenting the cases `for' and `against' their use. We then return to more traditional methods. Inspired by this, we present our own approach, applying ML within a `morally' justified framework. In particular, we use gradient descent on the Grassmannian manifold to identify an efficient subspace of sections for computation of the metric. This is further justified in a `Fourier modes' viewpoint, with conceptual similarities to the use of symmetries to restrict the basis. We test our algorithms on the Dwork family of quintics. Finally, we make some general comments about applications of ML in pure mathematics, whilst outlining avenues for future research.

\section{Mathematical background}
Following on from the general historical background, in this section we introduce the basic mathematical formalism necessary for understanding Calabi-Yau manifolds. We will assume knowledge of introductory differential geometry, as well as some familiarity with complex manifolds. Most of the material in this section is inspired by \cite{Cattani}, \cite{voisin_2002}, \cite{Huybrechts2004ComplexGA}. Some aspects of Riemannian geometry, with an emphasis on computation, will be covered in Section \ref{ref11} on Grassmannians.

We would also like for this paper to be somewhat accessible to computer scientists and mathematicians working primarily on computation and algorithms. It is highly likely that they would have much to contribute to solving the problems discussed here. We feel that the mathematical formalism outlined here can eventually be put partially to the side, once the concrete problem is formulated. Moreover, it is helpful to remember that 
 considered pointwise, many of the statements encountered are really just linear algebra. Amongst the confusion, it is also rather beautiful to see these structures interact.
\subsection{Complex and K\"ahler geometry}
An $n$-dimensional complex manifold is $2n$-dimensional real manifold $M$ equipped with an (integrable) complex structure $J: TM \rightarrow TM$, thought of as multiplication by $i$. A perhaps more intrinsic way of thinking about the space is local identification with $\mathbb{C}^n$, using holomorphic transition maps. It is surprising that this small change, $\mathbb{R} \rightarrow \mathbb{C}$, has far reaching consequences, leading to a  drastically new type of geometry. Following the usual course, a complex structure $J$ induces a decomposition of the complexified $TM \otimes \mathbb{C}$ into $\pm i$ eigenspaces, identified as the \emph{holomorphic} and \emph{antiholomorphic tangent spaces}, $T^{1,0}(M)$ and $T^{0,1}(M)$, respectively. One can then run through many definitions and theorems: of $(p,q)$-forms, the Hodge decomposition, and the Dolbeault operators inducing its own cohomology. Here we will skip this discussion, taking it as mostly assumed. The reader without this background is encouraged to consult \cite{Cattani}, \cite{Huybrechts2004ComplexGA}.

This setup allows us to consider K\"ahler manifolds. As already stated, perhaps the best introductory intuition to these is the existence of three compatible structures, the Riemannian, complex and symplectic. Starting with a complex manifold equipped with a bilinearly extended Riemannian metric, a natural class of metrics to consider is those that are compatible with $J$. We call such a triple $(M, J, g)$ a \emph{Hermitian manifold}. A basic and useful fact is the existence of a bijection between $J$-invariant metrics and \emph{Hermitian metrics} on the holomorphic tangent bundle, defined as a smoothly varying choice of sesquilinear form in each fibre. This comes naturally from the isomorphism between the real and holomorphic tangent bundles.

As is often the case in geometry, it is more convenient to study differential rather than bilinear forms. This leads to:
\begin{definition}[Fundamental 2-form]This is the two-form defined by $\omega(X,Y):=g(JX,Y)$. It is a real $(1,1)$-form, as can be checked.
\end{definition}

Crucially, the complex structure J, the fundamental 2-form, $J$-invariant Riemannian metric, and Hermitian metric on $T^{1,0}(M)$ all encode the same information. If the $J$ is not given, any two of these objects determine the other two. Now, we say the $(1,1)$-form $\omega$ is \emph{positive} if the corresponding Hermitian metric is. Of course, the corresponding $g$ is then also $J$-invariant and positive-definite. We will return to this notion frequently. 

\begin{definition}[K\"ahler manifold]
A Hermitian manifold such that d$\omega=0$ is called a K\"ahler manifold.
\end{definition}

The canonical example of a K\"ahler manifold is $\mathbb{P}^n$ equipped with the \emph{Fubini-Study (FS) metric}, given in local coordinates by: 
\begin{equation}
\omega_j := \frac{i}{2\pi} \partial \bar{\partial} \; \bigg( \frac{|z^0|^2+...+|z^n|^2}{|z^j|^2} \bigg),
\end{equation}
over the open sets $\{U_k: [z] \in \mathbb{P}^n: z_k \neq 0\}$. We will return to this expression, in modified forms, many times in this paper. One finally sees the connection to the symplectic viewpoint, since the closed fundamental $2$-form is a symplectic structure. 

In algebraic geometry, holomorphic line bundles and their sections are especially important, partly because there are no non-constant global holomorphic functions on compact, connected K\"ahler manifolds. They will be useful in this paper, playing a central role in Donaldson’s algorithm. A particularly important line bundle is the \emph{canonical bundle}, defined as the top degree exterior power bundle $\Lambda^{n,0}(M)$, and denoted by $K_M$. 

\begin{definition}[Calabi-Yau manifold]A compact K\"ahler manifold with trivial canonical bundle is called a Calabi-Yau.
\end{definition}

Here, we have taken the strongest possible definition; the adjunction formula \cite{Huybrechts2004ComplexGA} then makes it clear that every hypersurface in $\mathbb{P}^n$ defined by the vanishing locus of a homogeneous polynomial of degree $n+1$ is Calabi-Yau. For such a polynomial, it is easily calculated that there are $2n-1  \choose n-1$ adjustable coefficients. However, two such varieties are projectively equivalent if there exists a $GL(n, \mathbb{C})$ transformation between them, reducing the general parameters space by $n^2$. It turns out that the singularities lie on a codimension-one variety, so that the dimension of the space is just ${2n-1  \choose n-1}-n^2$. For a CY threefold in $\mathbb{P}^4$, this gives $101$ components. Later, we will focus on the one-parameter Dwork family embedded in $\mathbb{P}^{n-1}$, given by:
\begin{equation} \label{ref9}
\sum_{i=1}^{n} Z_i^n - \phi \prod_{i=1}^n Z_i =0, 
\end{equation}
for some fixed complex parameter $\phi$. 

For completeness, we provide some more elementary definitions and notation we will use in this paper, since these can vary in the literature. On Hermitian manifolds we can formulate the usual Levi-Civita connection, with the remarkable property that $\nabla J=0$ if and only if d$\omega=0$ (this gives new intuition for the K\"ahler condition). One can write down the usual curvature tensors, with many identities for the local components, easily referenced in the physics literature. A crucial feature of K\"ahler geometry is the local expression:
\begin{equation}
R_{i\bar{j}}= -\partial_i \partial_{\bar{j}}(\textnormal{log} \; \textnormal{det}(g_{k\bar{l}})),
\end{equation}
which gives the Ricci curvature purely in terms of a single real-valued function, the determinant of a matrix. It isn't really an exaggeration to state that this simple fact underlies a tremendous amount of what is interesting about K\"ahler manifolds, all the way up to and including Yau's results. The Ricci curvature can also be turned into a $(1,1)$-form to make it more amenable to geometric analysis, giving us the \emph{Ricci form}: $\rho(X,Y):=Ricci(JX,Y)$. 

Just as we can formulate the unique torsion-free metric-compatible Levi-Civita connection, for line bundles we can define the unique unitary and locally purely holomorphic connection with respect to a Hermitian metric, referred to as the \emph{Chern connection}. In this paper, the \emph{first Chern class} of a line bundle is then just the $\frac{i}{2\pi}$ multiplied onto the cohomology of the Chern curvature form. The latter can be calculated by the local formula:
\begin{equation}
\Omega_h = -\partial \bar{\partial}(\textnormal{log}|s|_h^2),
\end{equation}
where $s$ is any nonvanishing section. Now, we say a line bundle $\mathcal{L}$ is positive if $c_1(\mathcal{L})$ can be represented by a positive form. The first Chern class of a complex manifold $M$ is thus defined to be $-c_1(K_M)$, a purely topological quantity. Another obviously important line bundle is the \emph{hyperplane bundle} $\mathcal{O}(1)$, the dual to the tautological bundle $\mathcal{O}(-1)$, as well as its tensor powers, denoted by $\mathcal{O}(k)$. 

We can now state the foundational results discussed in the introduction.
\begin{theorem}[Calabi, 1957, \cite{Calabi}]
Given $(M,J, \omega)$, a compact K\"ahler manifold, and $\psi$, a real $(1,1)$-form representing $c_1(M)$, there exists a unique K\"ahler form $\tilde{\omega}$ such that $[\tilde{\omega}]=[\omega]$ and $\rho_{\tilde{\omega}}=2\pi \psi$.
\end{theorem}
\begin{theorem}[Yau, 1977, \cite{Yau}]
Given $(M,J, \omega)$, a compact K\"ahler manifold with $c_1(M)=0$, there exists a unique K\"ahler form $\tilde{\omega}$ such that $[\tilde{\omega}]=[\omega]$ and $\textnormal{Ric}(\tilde{\omega})=0$.
\end{theorem}
Since it can easily be shown that the curvature (of the Chern connection) of the canonical bundle, $\Omega_K$, satisfies $i\Omega_K=-\rho$ (where these objects are related by the usual identifications), one can finally see precisely by what was meant when we stated that Calabi had assumed the weakest topological condition for his conjecture. We see that on Calabi-Yaus, there is a unique Ricci-flat metric in every K\"ahler form cohomology class. Moreover, this is a vacuum solution to Einstein's equations.

Another characterisation of the Ricci-flat metrics that will be especially useful for numerical approximation is given by the \emph{Monge-Ampère (MA) equation}. A crucial feature of Calabi-Yaus, sometimes given as the definition, is the existence of a nowhere vanishing holomorphic $(n,0)$-form, often denoted by $\Omega$. Then we have:
\begin{theorem}Given a Calabi-Yau $(M,\omega)$, the K\"ahler form $\tilde{\omega}$ is Ricci-flat if and only if the Monge-Ampère equation: $\frac{\tilde{\omega}^n}{n!}= c \Omega \wedge \bar{\Omega}$, where $c$ is constant, holds.
\end{theorem}

From now on, we denote the K\"ahler volume form $\frac{\omega^n}{n!}$ by $\textnormal{d}\mu_g$, and the holomorphic volume form $(-i)^n\Omega\wedge \bar{\Omega}$ by $\textnormal{d}\mu_\Omega$. Also, we denote the corresponding volumes by $\textnormal{Vol}_g$ and $\textnormal{Vol}_\Omega$. Then, the MA equation gives an easy measure of deviation from Ricci-flatness:
\begin{equation} \label{ref13}
\mathcal{L}_{MA}= \frac{1}{\textnormal{Vol}_\Omega}\int_M \left\lvert 1-\frac{1}{\kappa}\frac{\textnormal{d}\mu_g}{\textnormal{d}\mu_\Omega} \right\lvert \; \textnormal{d}\mu_\Omega,
\end{equation}
where $\kappa$ can either be the ratio of volumes or set to a constant. In the latter case, this controls the K\"ahler class. One could also use integrals of scalar-valued quantities associated to the curvature, but this tends to involve the costly computation of derivatives \cite{moduli}.

Finally, we include some useful theorems about K\"ahler manifolds, and general remarks that will be helpful for the remainder of the paper. We have mentioned the importance of global holomorphic sections. A main reason for their significance is that they allow embeddings into higher-dimensional projective spaces. In the literature, it is often stated that these sections are homogeneous monomials of degree $k$ in the homogeneous coordinates. This can be confusing, since strictly speaking such monomials are not well-defined as functions. The correct interpretation is to divide by the local trivialisations $(z^i)^k$ over $U_k$, so that everything transforms correctly. 

The embeddings will be the standard choice in algebraic geometry, given by:
\begin{equation}
    i: M \rightarrow \mathbb{P}^N, \; \; i(z):=[s^0(z):...:s^N(z)],
\end{equation}
where $\{s_0,...,s_N\}$ are the $N+1$ global sections of a line bundle $\mathcal{L}$.  An iconic result of Kodaira states that for compact $M$, the bundle $\mathcal{L}$ is \emph{ample} (has enough sections to embed for a sufficiently high power $\mathcal{L}^k$) if and only if it is positive. For the cases we will consider, $\mathcal{L}$ will always be very ample, which means that it already admits enough sections.

Given a K\"ahler manifold $X$, the $\overline{\partial}$-Poincaré lemma (a simple amendment of the usual result, with restricted domains \cite{voisin_2002}) implies an elementary but useful result:
\begin{lemma}Let $(M,\omega,g)$ be a K\"ahler manifold. Around any $p \in M$, there exists an open set $U$ and a real function $v \in C^\infty(U)$, such that $\omega= i \partial \bar{\partial} v$, or equivalently, $g_{i \bar{j}}= \frac{\partial^2 v}{\partial z^i \partial \bar{z}^{\bar{j}}}$.
\end{lemma}

Put simply, $\omega$ can be locally represented by a smooth function, known as the potential. The problem of Ricci-flat metrics reduces to finding a collection of functions $v$ with the right properties. However, it isn't a priori easy to produce functions giving K\"ahler metrics. In particular, the positivity of the Riemannian metric $g$ is hard to enforce, apart from the case of Fubini-Study (FS). For the FS metric, the form $\omega_{FS}$ on $\mathbb{P}^n$ is uniquely invariant under the action of $U(n+1)$. This means it suffices to show positivity at the origin, then restrict to the variety. 

\subsection{Algebraic metrics}
In this subsection we introduce the K\"ahler \emph{algebraic metric} ansatz, and show that it can be interpreted as a truncated Fourier-like series.

Consider a projective variety $X$ equipped with the ample line bundle $\mathcal{O}(k)$. Let  $\{Z^0,... \;,Z^{N_k}\}$ denote the set of homogeneous monomials of degree $k$ in homogeneous coordinates. The eigenfunctions of FS Laplacian, expressible as: $\psi^{\alpha \bar{\beta}}=\frac{Z^\alpha \bar{Z}^{\bar{\beta}}}{(\sum_i |z^i|^2)^k}$, form a complete orthonormal basis for functions in $\mathbb{P}^n$ \cite{Donaldson}. Therefore any function in $C^\infty(X)$ can be approximated by an expansion in the $\psi^{\alpha \overline{\beta}}$ (restricted to $X$, with some potential analytic subtleties). This can be used to produce Ricci-flat metrics.

A convenient general fact is the following. Consider a holomorphic vector bundle $E \rightarrow M$. Any two hermitian metrics on $E$, $h_1$ and $h_2$, are related by a positive rescaling, so that $\langle.,.\rangle_{h_1} = e^{-f} \langle.,.\rangle_{h_2}$, for a function $f:M \rightarrow \mathbb{R}$. Given $s \in \Gamma(E)$, it follows that the difference: $\textnormal{log}(|s|^2_{h_1}) -\textnormal{log}(|s|^2_{h_2})$ is a global real function on $M$.

Let us return to the K\"ahler case. Embed $\mathbb{P}^n$ into higher-dimensional projective space using $\mathcal{O}(k)$. Then pull back the FS metric associated to a positive-definite hermitian $h_{\alpha \bar{\beta}}$, to the variety $X$. This gives local potentials of the form:
\begin{equation} \label{ref1}
K_j([z])= \frac{1}{ \pi k} \textnormal{log} \; \frac{Z^\alpha h_{a\bar{\beta}}\bar{Z}^{\bar{\beta}}}{|z^j|^{2k}},
\end{equation}
over the open set $U_j$. The $\frac{1}{k}$ prefactor ensures that the K\"ahler class remains the same as the FS metric on $X$, for all $k$. Crucially, pulling back preserves positivity; we know that this is otherwise hard to ensure. Following \cite{Donaldson}, \cite{headrick}, we call this the algebraic metric ansatz.

Another way of interpreting the local potentials \ref{ref1}, inside the logarithms, is as hermitian metrics on $\mathcal{O}(k)$, using the local sections $(z^j)^k$ to trivialise. These expressions satisfy the correct transformation rules with the transition functions $g_{\alpha \beta}= \frac{(z^\alpha)^k}{(z^\beta)^k}$. By the above reasoning, the difference $K-K_{FS}$ is a global real function, where $K_{FS}$ denotes the FS potential $\frac{1}{\pi}\textnormal{log}({\sum_i |z^i|^2)}$. So it can be approximated by a truncated expansion in the $\psi^{\alpha \overline{\beta}}$. By direct calculation:
\begin{equation} \label{ref8}
K - K_{FS} = \frac{1}{\pi k}\textnormal{log} \; (Z^\alpha h_{\alpha \bar{\beta}} \bar{Z}^{\bar{\beta}}) - \frac{1}{\pi} \textnormal{log} \; (\sum_i |z^i|^2) = \frac{1}{\pi k} \textnormal{log}  (\psi^{\alpha \bar{\beta}} h_{\alpha \bar{\beta}}),
\end{equation}
one finds that the elements $h_{\alpha \overline{\beta}}$ are the coefficients of the eigenfunctions \cite{headrick}. Having established this, we seek a matrix such that the potential $K$ approximates the unique Ricci-flat representative of the FS K\"ahler class. Since the dimension $N_k$ grows like $O(k^n)$, for computational efficiency, it may make sense to use a different truncated basis.
\section{Existing methods}
The universal approximation theorem \cite{HORNIK1989359} implies that neural networks are function approximators to an arbitrary degree of accuracy. In this section, we review ML approximations and some simple statistical background, leading naturally into a summary of older methods.
\subsection{Neural network approaches} \label{sec3.1}
A neural network (NN) can be used as a form of ansatz for either the metric or the potential. The NN outputs either a matrix-valued function $g_{NN}$ or a smooth function $\phi_{NN}: M \rightarrow \mathbb{C}$, learning the metric as $\omega_{FS}+\partial \bar{\partial} \phi_{NN}$. At a point $p\in M$, the non-zero components are an $n \times n$ Hermitian matrix $h_{\alpha \bar{\beta}}$. Recall that the positive definiteness of $g$ is equivalent to the positive-definiteness of this matrix, which can be interpreted as a Hermitian metric on $T^{1,0}(M)$.

After inputting the homogeneous coordinates $\{z_1,...,z_n\}$, or their real and imaginary parts, the neural network ansatz is a composition of affine transformations and non-linear activations. In other words, $\textbf{s}_{output}= L^n \circ \sigma^n \circ L^{n-1} \circ ... \circ \sigma^1\circ L^1(\textbf{s}_{input})$, where the affine transformations $L^i$ are multiplication and addition by a (real or complex) weight matrix $\textbf{W}^i$ and a (real or complex) bias vector $\textbf{b}^i$, and the $\sigma^i$ the nonlinear activation functions. Training, in this context, means inputting a collection of uniformly sampled  points, then minimising the loss $\mathcal{L}$ by optimising the weights and biases. 

Thus far, most papers - for example  \cite{moduli}, \cite{challenger}, \cite{Larfors:2022nep}, \cite{berglund2023machine} - have used the same overall idea, taking an $\mathcal{L}$ of the form:
\begin{equation} \label{ref2}
\mathcal{L}= \lambda_1 \mathcal{L}_{MA}+\lambda_2 \mathcal{L}_{Transition}+\lambda_3 \mathcal{L}_{dJ}(+\mathcal{L}_{Class}),
\end{equation}
where the terms should be understood as numerical integrals on $M$, calculated by a Monte Carlo method.
The constants $\lambda_i$ control the contribution each term have on the overall loss. Depending on the ansatz, some may be set $0$, as the constraints are automatically satisfied. We briefly consider the purpose of each of them.

The first term is just the same as in \ref{ref13}. The second term enforces patch agreement on overlaps, for example:
\begin{equation}
\mathcal{L}_{Transition}= \sum_{i \neq j} \int_{U_i \cap U_j} |\phi_i-\phi_j| \; \textnormal{d}\mu_\Omega.
\end{equation} \label{ref10}
This is necessary if learning takes place on several different patches, with each network outputting $\phi_i$ on $U_i$. If the NN outputs $g_{NN}$, \ref{ref10} is replaced by the condition that it transforms correctly under the usual transition functions. Finally, the third term enforces closedness of the associated $(1,1)$-form $\omega$:
\begin{equation}
\mathcal{L}_{dJ}= \mathlarger{\mathlarger{\sum}_{ijk} \left\lvert \textnormal{Re}\bigg( \frac{\partial g_{i\bar{j}}}{\partial z_{k}}-\frac{\partial g_{k\bar{j}}}{\partial z_{i}}\bigg)\right\vert}_2 + \left\lvert \textnormal{Im}\bigg( \frac{\partial g_{i\bar{j}}}{\partial z_{k}}-\frac{\partial g_{k\bar{j}}}{\partial z_{i}}\bigg)\right\vert_2,
\end{equation}
where $|.|_2$ denotes the $L_2$-norm on $M$. Note that the simplicity of this condition (only the $\partial$ term is vanishing) follows from the reality of $\omega$.

Depending on the setting, one can include the term $\mathcal{L}_{Class}$ to enforce a choice of K\"ahler class. If $h^{1,1}=1$, this is fixed by volume, controlled already by $\mathcal{L}_{MA}$. To learn a form cohomologous to Fubini-Study, one expresses $\omega_{FS}$ as a linear combination of a basis for $H^{1,1}(M)$, i.e. $\omega_{FS} = t^i \omega_i$, then calculates the corresponding volume with known intersection numbers. The deviation of $[\omega_{NN}^2]$ from $[\omega_{FS}^2]$ can then be penalised by adding an appropriate term, of the form $\mathcal{L}_{class} \propto|\int_M \omega^2 \wedge \omega_{i} \; \textnormal{d}_{\Omega}- \textnormal{\emph{Sum of intersection numbers}}\;|$, summed over the basis elements (these volumes are known as the slopes of the line bundles $\mathcal{L}$ with $c_1(\mathcal{L})$ given by $\omega_i$ \cite{Larfors:2022nep}). We note that this only enforces that $[\omega_{NN}^2]=[\omega_{FS}^2]$, which depending on the structure of $H^2(M)$, may not be equivalent to $[\omega_{NN}]=[\omega_{FS}]$. For example, in the case of $h^{1,1}=2$, if there exist nilpotent elements in the cohomology linearly independent from $\omega_{FS}$, it is easy to come with examples $[\omega_{FS}+\mu]$ such that $[\omega_{FS}+\mu]^2=[\omega_{FS}]^2$, with $\mu \neq 0$. Thus it is unclear to us whether this condition will enforce the right K\"ahler class, beyond simple situations like the Fermat quintic with $h^{1,1}=1$.

Some have found the most numerically stable results have been achieved by initialising with $g_{FS}$, the Fubini-Study metric \cite{berglund2023machine}. Others initialise with standard Gaussians \cite{Larfors:2022nep}. The NN can learn a variety of different relations between $g_{NN}$ and the desired Ricci-flat $g_{RF}$, for example $g_{RF}=g_{NN}$ or $g_{RF}=g_{FS}+g_{FS}\cdot g_{NN}$, making this approach very flexible to the desired application. Of course, the  great strength of neural networks is their ability to learn functions to a high degree of accuracy in relatively quick times, as confirmed by Monge-Amperè errors of $O(10^{-3})$, (or better) being achieved \cite{berglund2023machine}, \cite{butbaia2024physicalyukawacouplingsheterotic}. Moreover, partial derivatives of the resulting metrics are easily computed, in contrast, for example, to solving the equation on a lattice. This means that in the case of the standard embedding, the harmonic tangent bundle-valued $(0,1)$-forms are easily computed. One can parameterise the appropriate exact corrections to the cohomology basis elements using another neural network $s(\theta)$, then minimise the relevant integrals over $\theta$ with respect to the volume form inherited from $g_{NN}$. This is a crucial step in the computation of the physically-meaningful normalised Yukawa couplings, and is easily incorporated in a NN pipeline. 

Beyond the speed of computation and relative flexibility, another key advantage of the neural network ansatz is that it does not rely on the discrete symmetries of the CY itself, for example interchange of the homogeneous variables or multiplication by roots of unity. The networks achieve good results without directly implementing such symmetries on the network, in contrast with more traditional methods, which require a manual reduction of the relevant basis for computational efficiency. This once again suggests a synthesis of approaches, which we will return to in Section \ref{ref14}.

\subsection{Metric positivity} \label{sec1}
Machine learned metrics have enabled significant strides in the realm of string phenomenology. However, it is worthwhile considering the applicability of neural networks in this area of research.

For one thing, the geometry of the problem seems to enforce certain architecture choice. These have not yet widely adopted in the literature. In our view, if the metric is being learnt directly, it is essential that one uses the local decomposition $g_{NN}=LDL^\dag$. Here, $D$ is a positive diagonal real matrix of eigenvalues and $L$ is an invertible lower triangular matrix with ones along the diagonal. What this means is that the outputted $g$ is manifestly positive (this approach was taken in \cite{moduli} and \cite{challenger}, but not in \cite{Larfors:2022nep} and \cite{berglund2023machine}. The matrices $D$ and $L$ can be outputted by two different, simultaneously optimised, networks. However, local agreement and closedness do not automatically hold, so $\lambda_2$ and $\lambda_3$ in \ref{ref2} must be taken as non-zero. If the metric is learned directly as a $3\times 3$ tensor on a CY threefold, without any such decomposition, positive eigenvalues are not guaranteed.

A potentially cleaner approach is to directly learn a functional correction to $\omega_{FS}$; this has become the standard method in the literature. On a compact CY, by the $\partial \bar{\partial}$-lemma, there is a global real $\phi$ such that $\omega_{RF}=\omega_{FS}+ i {\partial} \bar{\partial} \phi$, unique up to an additive constant, which can be approximated by a NN. In this case $\omega_{NN}$ is a true global and closed form by construction. Unfortunately, the positivity of the metric is again no longer guaranteed. The relevant space of functions can be denoted by:
\begin{equation}
\mathcal{H}=\{\phi \in C^\infty(M) : \omega_\phi = \omega_0+i\partial \bar{\partial} \phi >0 \}.
\end{equation}
Obviously, this is a tricky condition to satisfy, and is required to state the Calabi conjecture, as noted by Yau himself \cite{Yau2008ASO}. To our knowledge, there are no other simple mathematical characterisations of this class, and it is unclear how it can be enforced, even weakly, on the outputs. Surprisingly, we found the incorrect statement $\omega_{FS}+i \partial \bar{\partial} \phi$ is automatically K\"ahler several times in the literature \cite{Larfors:2022nep}. We believe that existing consistency checks - for example in the slope or volume calculations of \cite{Larfors:2022nep} - only address cohomological information, for example, that $[\omega_{FS}^3]=[\omega_{NN}^3$]. This does not say anything about positivity. 

One justification for overlooking this issue in the literature is that the $\mathcal{L}_{MA}$ loss and $\omega_{FS}$ initialisation somehow encourages the NN to learn positive forms. However, we find this argument unsatisfying. Current ML methods hope that the network learns positive forms on uncountably many tangent spaces on the manifold, an intrinsically difficult problem. Note that if the network is initialised as positive definite, it is plausible that the Monge-Ampère loss will discourage $\textnormal{det}(g_{\alpha \bar{\beta}})$ from crossing $0$, since the loss would become singular on some points. However, the loss is an integral, and functions with singularities can obviously still be integrable. We believe that this poses a serious mathematical problem to learning a functional correction to the potential.

It is also very unlikely that this condition can be checked by pointwise sampling. Consider a simple semi-definite example, e.g. the $(0,2)$ tensor: $
T= \sum_i (t_i-\alpha_i)^2\; \textnormal{d}t_i ^2$
on $\mathbb{R}^n$. Sampling points and checking the eigenvalues would always give positive answers, away from the measure zero hypersurfaces $t_i=\alpha_i$. This undermines the reliability of the consistency checks performed, for example, in the appendices of \cite{berglund2023machine}. More generally, on an open $U$-trivialisation, a semi-definite function matrix $M$ will fail to be positive definite on the measure-zero set $\{ p \in U \; | \; \textnormal{det}(M|_p)=0 \}$. If the output of the NN is real analytic, the identity theorem implies that the points of vanishing are isolated \cite{lang1985complex}. 

Informally, there seems to be a `no free lunch principle' for the ML-derived metrics. In practice, anytime one of the $\lambda_i$ can be zero by a choice of ansatz, the other $\lambda_i$ conditions become harder to enforce. This underlines the delicate nature of K\"ahler geometry and motivates a reexamination of older methods. In particular, the algebraic ansatz \ref{ref1} is K\"ahler by construction. 

\subsection{Problems with error composition}
Using weighting parameters such as the $\lambda_i$ in Eq. \ref{ref2} is a common way of combining several different objectives in machine learning. The classic example is an underdetermined regression task, where additional error terms relating to the regression coefficients are introduced to make the problem well-posed (this is known as Tikhonov or ridge regression). However, the setting we are interested in here is significantly different. The functions in $\mathcal{L}$ are not soft constraints on the problem but hard theoretical requirements, which cannot be `approximately' satisfied in any obvious sense. From a statistical learning perspective, a `soft constraints' approach allows an expression of prior beliefs about the model parameters, so that in the aforementioned example, $L^2$ regularisation is equivalent to a Gaussian prior \cite{bishop}. We briefly review this perspective now. 

Consider a supervised learning problem on a dataset $\mathcal{D}$ of $N$ labelled data points $(\textbf{x}_i, \textbf{y}_i)$. We seek a $\textbf{w}$-parameterised function $f(\textbf{w},.)$, such that $f(\textbf{w},\textbf{x})\approx \textbf{y}_i$, for all $i$. Having selected objective function $L$, e.g. the mean squared error over $\mathcal{D}$, we are given a set of $M$ constraint functions $C_j(.)$. Following \cite{márquezneila2017imposinghardconstraintsdeep}, one can think of two different problems: \begin{subequations} \label{ref4}
\begin{gather}
   \textbf{w}^*= \underset{\textbf{w}}{\textnormal{argmin}} \; L(\textbf{w}), \; \; \textnormal{s.t.} \; \;C_j(f(\textbf{w},\textbf{x}_i))=0, \; \; \textnormal{for} \; 1 \leq i \leq N, 1 \leq j \leq M,\label{ref4a}  \\
    \widetilde{\textbf{w}}^* = \underset{\textbf{w}}{\textnormal{argmin}} \; \big(L(\textbf{w})+\sum_{i, j} \lambda_j C_j\big( f(\textbf{w}, \textbf{x}_i)\big)^2 \big), \label{ref4b} 
\end{gather}
\end{subequations}
where in \ref{ref4a}, \emph{hard constraints} are enforced, and in \ref{ref4b}, soft constraints are enforced. It is worthwhile considering the different formulations, in particular that \ref{ref4b} necessarily involves a compromise between minimising $L$ and satisfying the $C_j(.)$. 

Of course, the problem with case \ref{ref4a} is that it becomes very hard to solve directly, when the $\textbf{w}$-space becomes high-dimensional, or $\mathcal{D}$ gets sufficiently large. Things must be simplified, essentially by relaxing the hard constraints. Alternatively, one can try to enforce them by a clever choice of architecture, but this may not be straightforward. An example of the former is \cite{márquezneila2017imposinghardconstraintsdeep} who find the solution to the appropriate Lagrangian linear system with a Krylov method; an example of the latter is \cite{agrawal2019differentiable}, who incorporate a differentiable projection operator onto the constraint space, into the gradient descent. 

Perhaps more seriously, the issue with case \ref{ref4b} is that $C_j(.)$ may be poorly enforced. The limitations of NNs as cheap `approximate solvers' is widely accepted in the ML literature \cite{márquezneila2017imposinghardconstraintsdeep}, \cite{dc3}, \cite{bengio}. The issue of infeasible solutions has been discussed in the contexts of power grids, climate models, and many-body physics, amongst others \cite{opf}, \cite{beucler2019achievingconservationenergyneural}, \cite{NEURIPS2018_842424a1}. In all these situations, solutions must conform to known physical laws, and deviations can have serious practical consequences, for example power outages. In our view, K\"ahler-Einstein manifolds fall in this paradigm.

We believe that soft constraints should always be thought of as imposing some form of prior on the model parameters \cite{bishop}. The practical justification is that it prevents overfitting, meaning better generalisation to unseen data. However, the addition of terms to the loss function affects the loss landscape (in both convex and nonconvex cases) leading, for example, to getting trapped in small $\textbf{w}$ local minima \cite{exact}, \cite{NIPS2017_1abb1e1e}. There are results showing the genericity of bad local points under the assumption of weight decay regularisation \cite{exact}, \cite{ding2020suboptimallocalminimaexist}. For the calculation of K\"ahler metrics many more loss terms are typically being added, almost certainly making the issue more serious.
\subsection{Approximate K\"ahler metrics?} \label{sec2}
In \ref{ref2}, $\mathcal{L}_{MA}$ is the objective function whilst $\mathcal{L}_{Transition}$, $\mathcal{L}_{dJ}$ and $\mathcal{L}_{Class}$ can be thought of as constraints. Whilst regularisation implies that bad local minima exist, other results suggest that for sufficiently wide neural networks, there may be few such bad points, in a sense that can be made precise \cite{lecun}, \cite{nguyen}. The behaviour of NNs during training has been studied, for example using Neural Tangent Kernels (NTKs) \cite{jacot2018neural}. These can be used to analyse convergence of NNs in function space in an infinite width limit. For our purposes, it is sufficient to note that even the solutions lying near a local minima which is very `good' in terms of error are quite simply \emph{not K\"ahler metrics}. Thus Yau's existence and uniqueness proofs do not make sense for them. 

The basins of attraction of local minima can be escaped by introducing stochasticity, an example being simulated annealing. However, in much of the existing work, the $\mathcal{L}_{dJ}$ and $\mathcal{L}_{Transition}$ errors appear to be converging to small, positive fixed values, even as $\mathcal{L}_{MA}$ continues to decrease. Most pathological are the cases when loss terms increase during training, following an initial dip \cite{Press2007} (for example, Figure 5 of \cite{moduli} or Figure 10 of \cite{challenger}). From the perspective of regularisation, we would like a $\delta$-function prior on K\"ahlericity, but it is not clear how to impose this on a NN architecture. Further, it is unclear how gradient information outside of the range of feasibility should be interpreted. Morever, this is all ignoring the more serious positivity issue, outlined already in Section \ref{sec3.1}.

For now, we pass over further discussion and interpretation of the NN metrics. We believe that a strong case has been made for development of alternative approaches, in particular those that are based on principled ansatzes. In the next section, we review Donaldson's algorithm, which forms the basic framework for our method, augmented with vanilla gradient descent. 

\subsection{Donaldson's algorithm, in brief} \label{sec3}
One would like to inform any learning by the geometry underlying the problem. This brings us back to algebraic potentials, and the problem of finding the right matrices $h_{\alpha \overline{\beta}}$. The first steps in this direction were taken in the pioneering work of Donaldson \cite{Donaldson}, using a sequence of embeddings to approximate the potential \cite{Scalar}, \cite{Scalarii}, \cite{tian}. In its simplest form, Donaldson's algorithm can be understood in terms of two limits.
\begin{enumerate}
    \item One is for a fixed line bundle $\mathcal{O}(k)$.
    \item The second is for the bundles $\mathcal{O}(k)$, as $k \rightarrow \infty$.
\end{enumerate}
Given a positive-definite hermitian $h_{\alpha \bar{\beta}}$, define the integral operator:

\begin{equation} \label{ref5}
(T(h))_{\gamma \bar{\delta}} := R \int_{M} \frac{Z^\gamma \bar{Z}^{\bar{\delta}}}{\sum (h^{-1})_{\alpha \bar{\beta}} Z^\alpha \bar{Z}^{\bar{\beta}}} \textnormal{d} \mu_\Omega,
\end{equation}
where $d \mu_\Omega= \Omega \wedge \bar{\Omega}$, as before. Then we have the following:
\begin{theorem} \label{ref16} Consider a compact CY embedded in $\mathbb{P}^n$. For each line bundle $\mathcal{O}(k)$, there exists a unique fixed point of the T-operator up to scale, and iterating $T(h_0)$ for any initial $h_0$ converges to it. Moreover, the sequence of corresponding balanced K\"ahler forms $\omega_\infty^k$ converges to the Ricci-flat metric as $k\rightarrow\infty$.
\end{theorem}
We call a fixed point of the $T$-operator, for a given bundle $\mathcal{O}(k)$, the \emph{balanced} metric. Thus Theorem \ref{ref16} can be rephrased as stating the balanced metrics converge to the Ricci-flat one, in the limit that $k \rightarrow \infty$. One show that convergence takes places with a decay in error that is $O(k^{-2})$ \cite{Donaldson}. Thus we have an algorithmic method for tuning the coefficients occurring in the algebraic potential \ref{ref1} in terms of a simple integral operator.

However, it was found to have one major flaw: the curse of dimensionality associated to the growth of the space of global sections $\{Z^0, ... , Z^{N_k}\}$. It turns out that the number of necessary operations scales like $O(k^{4n})$, where the biggest bottleneck is the $O(N_k^2)$ points necessary for the numerically stable computation of \ref{ref5}.  Beyond $k=10$, the calculations become far too much for the capabilities of a typical laptop. For physical applications, where one hopes to move in the moduli space and compute associated metrics rapidly, these timings are unacceptable, even using powerful computing. It is therefore natural to consider methods which maintain K\"ahlericity, but converge more quickly in $k$. 
\subsection{Energy functionals} \label{sec3.6}
Although they converge in the limit, the balanced metrics are not directly optimising for Ricci-flatness \cite{Donaldson}, \cite{headrick}. Donaldson predicted a sequence of `refined' algebraic metrics, such that the $\sigma$-error decays exponentially with $k$. Following this suggestion, Headrick and Nassar \cite{headrick} used energy functional minimisation to achieve this. They considered the integral of a convex, differentiable, bounded below $F(\eta)$.  As noted by many classic references, it is generally accepted in numerical analyis that such an approach is easier than directly solving a PDE \cite{Press2007}. We review their method now. 

Let us assume, without a loss of generality, that $F(\eta)$ attains a unique minimum at $\eta=1$. The aim is to extremise the functional:
\begin{equation} \label{ref17}
E_F[\omega]:=\int_X F(\eta) \; \textnormal{d}\mu_\Omega,
\end{equation}
for variations of $\omega$ within the algebraic class. Thus one can derive the first order variation in \ref{ref17}, corresponding to $\omega \rightarrow \omega+i\partial \bar{\partial} \phi$, which gives the appropriate Euler-Lagrange (E-L) equations. Note $\phi$ will take a particular `algebraic' form. 

We specialise to the case that $F=(\eta-1)^2$, justified as the leading order contribution whenever $\eta \approx1$. This gives the E-L equations:
\begin{equation} \label{ref18}
\int_X \eta  \nabla_\omega^2 \; \eta \frac{Z^\gamma \bar{Z}^{\bar{\delta}}}{Z^\alpha h_{\alpha \bar{\beta}} \bar{Z}^{\bar{\beta}}} \; \textnormal{d}\mu_\Omega=0, \; \; \textnormal{for} \; \gamma, \delta = 1,...,N_k.
\end{equation}
The function $\nabla_\omega^2 \eta$ is integrated against the eigenstates of the FS Laplacian, so that \ref{ref18} becomes a Galerkin-like condition for $\nabla_\omega^2 \eta=0$ with these orthogonal functions. This implies the constancy of $\eta$, equivalent to solving Monge-Ampère. To enforce these equations, Headrick and Nassar use a sequence of algebraic manipulations and a Levenberg-Marquardt method. The result is a more direct method for imposing Ricci-flatness in the coefficients of the potential.

This approach can still be considered as the state-of-art in many respects. In our personal view, it is unclear why neural networks have recently been so strongly preferred in the literature. Because the error decays exponentially with $k$, one can achieve  errors at the order of $10^{-3}$ by $k=8$. Like Donaldson's algorithm, the Headrick and Nassar approach the advantage of manifest K\"ahlericity. It has also been combined with ML, used to speed up the functional minimisation \cite{calmon}. 

However, such an approach still suffers from the curse of dimensionality. Headrick and Nassar only implement on CYs with a great deal of symmetry, such as the Fermat quartics and quintics. They use symmetries of the defining polynomial, for example $z^i$ interchange, to greatly restrict the class of polynomials occurring in the potential. Thus a method relying less on naive symmetries would be useful for more general calculations.

Moreover, the gradient approach taken in \cite{headrick} is just one possibility, involving a specific choice of optimisation procedure. For example, stochastic gradient descent approach on the $h$-matrix was taken in \cite{moduli}. We were not sure how the matrix was enforced to be positive Hermitian in either case. Since the problem has been reduced to a single potential matrix, it is much easier than for the NNs. Nevertheless, a gradient descent approach that searches strictly on the objects of interest, i.e. the manifold of positive Hermitian matrices, seems desirable, avoiding a potentially problematic `projection step' in the updates.
\subsection{No free lunch}
Previously we formulated a `no free lunch' principle for the ML metrics. This stated that whenever the NN directly enforced some desired properties, the other conditions became harder to satisfy. Now it can be generalised to Donaldson's algorithm and the energy functional method. In the first case, one `pays' for manifest K\"ahlericity and a convergence guarantee with high computational cost. In the second case, one `pays' for K\"ahlericity and exponential convergence by restricting the set of CYs one can implement on. From that perspective, an approach which achieves the right compromise between our different requirements is to be aimed for, seeing as a perfect solution does not seem possible. 

\section{Donaldson's algorithm}
Having addressed other methods, we return to Donaldson, with a greater focus on the mathematical formulation. By carefully defining `balancedness', it becomes clear that it can be restricted to a smaller space. This fact will lead naturally into our proposal. 
\subsection{Balanced metrics}

 Given a $d$-dimensional Calabi-Yau $X$ equipped with a positive Hermitian bundle $\mathcal{L}$, the method is based around two kinds of data, uniquely determining each other. These are:
 \begin{enumerate}
 \item Hermitian forms on the space of global sections of $\mathcal{L}^{\otimes k}$, denoted by $H^0(X, \mathcal{L}^k)$, 
 \item
Hermitian metrics on the line bundles $\mathcal{L}^{\otimes k}$. 
\end{enumerate}
Firstly, every Hermitian line bundle metric $(.,.)_h$ can easily be turned into a Hermitian form on $H^0(X, \mathcal{L}^k)$ by taking an $L^2$ inner product:
\begin{equation} \label{ref6}
\langle s^\alpha, s^\beta\rangle_{\textnormal{Hilb}(h)}:= \frac{N_k}{\textnormal{Vol}_{\textnormal{d}\nu}}\int_{\mathbb{P}^n}  (s^\alpha, s^\beta)_h \; \textnormal{d}\nu,
\end{equation}
where $\textnormal{d}\nu$ is a \emph{positive Radon measure} \cite{folland2013real}. 

Conversely, given a Hermitian form on $H^0(X,\mathcal{L}^k)$, say $\langle.,.\rangle_G$, there is a Hermitian metric on $\mathcal{L}^{\otimes k}$, FS($G$), uniquely characterised by the pointwise condition:
\begin{equation} \label{ref7}
\sum_i |t^i|^2_{\textnormal{FS}(G)}=1,
\end{equation}
where $t_i$ is an orthonormal basis for $H^0(X, \mathcal{L}^k)$ with respect to $G$. Importantly, this does not depend on the choice of basis, provided it is orthonormal, nor does it depend on a rescaling of $h$.

Now, the $h_{\alpha \bar{\beta}}$ matrix occurring in the algebraic potential \ref{ref1}, can be interpreted as line bundle metric in the following sense:
\begin{equation}
(s^\alpha, s^\beta)_h = \frac{s^\alpha \cdot \bar{s}^\beta}{\sum h_{\gamma \bar{\delta}} s^\gamma s^{\bar{\delta}}}.
\end{equation}
 It turns out, following a calculation, that if we apply \ref{ref7} to a form on $H^0(X, \mathcal{L}^k)$, encoded in a positive-definite Hermitian matrix $G_{\alpha \bar{\beta}}=\langle s^\alpha, s^{\bar{\beta}} \rangle$, we get the line bundle metric:
\begin{equation}
(s^\alpha, s^\beta)_{FS(G)} = \frac{s^\alpha \cdot \bar{s}^\beta}{\sum (G^{-1})_{\gamma \bar{\delta}} s^\gamma s^{\bar{\delta}}},
\end{equation}
using the inverse matrix. To summarise: we have two maps, Hilb and FS, which map from line bundle metrics to Hermitian forms on $H^0(X, \mathcal{L}^k)$, and vice versa. Composing \ref{ref7} and \ref{ref6}, one recovers exactly the T-operator already given in \ref{ref5}. This is the true definition.

\begin{definition} \label{ref35} A Hermitian metric $h$ is $\nu$-\emph{balanced}, if  $T(h):=\textnormal{FS(Hilb}(h))$.\end{definition} 
Since a bundle metric $h$ determines a Hermitian form, and vice versa, one can think of $G$, $h$, or the pair $(G,h)$ as each being itself `balanced'. Then an embedding $i_k$ into $\mathbb{P}^{N_k-1}$ with a basis of sections orthonormal with respect to such a $G$ is called a \emph{balanced embedding}. The pullback K\"ahler $\omega_k := \frac{1}{k}i_k^*(\omega_{FS})$ is also called a balanced metric. Moreover, we may call a pair $(X,\mathcal{L})$ balanced, if a balanced embedding exists. 

\subsection{Convergence results}
Having defined the balanced metrics, it remains to show what we ultimately want: that they converge to Ricci-flat representatives. 
\begin{theorem} \label{ref20}
Given $\textnormal{d}\nu$, a positive Radon measure on $\mathbb{P}^n$, there is a unique $\nu$-balanced pair $(h,G)$ up to scaling. Moreover, if $h_\infty$ is the balanced line metric, then $ T^r(h_0)$ converges to $h_\infty$ as $r \rightarrow \infty$.
\end{theorem}

\begin{theorem} \label{ref19}
The sequence of a balanced K\"ahler metrics $\frac{1}{k} \omega_k$ converges to the metric with corresponding volume $\textnormal{d}\nu$, as $k \rightarrow \infty$.
\end{theorem}
These results have generalised Theorem \ref{ref16} to a greater class of measures. Of course taking the Radon measure given by the top degree volume form, and restricting to $X$, gives an algorithmic procedure for generating approximations Ricci-flat metrics on Calabi-Yaus. The definition of the Radon measure means that this is robust to integral approximations by sums of point masses. As an aside, it is an interesting mathematical fact that no analytic expressions for balanced metrics have yet been found. 

Because it will in some sense relate to our own method, we have included our own rough outline of a proof to Theorem \ref{ref19} in Appendix \ref{ref21}. Although it is not novel, we use ideas from Local Index Theory, which we have not yet come across in the literature in this area. The proof to Theorem \ref{ref20} is more standard and can be found in \cite{Donaldson} and \cite{sano} (for two different integration measures - the case of interest to us is covered in the former).

\subsection{Summary}
To recap, Donaldson's algorithm is a simultaneously global and local approach. It is a compelling geometric method that unfortunately suffers from the curse of dimensionality. However, it is natural to consider whether aspects of it can be coupled with a more proabilistic/ML framework. In particular, we note that the $O(N_k^2)$ growth in computational cost is associated to the dimensionality of $H^0(X,\mathcal{L}^k)$. If a notion of `balancedness' could be sensibly formulated on a subspace, this large cost could be reduced. The mathematical way of thinking systematically about subspaces is the Grassmannian, which we now turn to. 

\section{Grassmannians} \label{ref11}
In theoretical physics, the positive Grassmannian has been connected to scattering amplitudes in  $\mathcal{N}=4$ super Yang-Mills theory \cite{arkani2012scattering}. In machine learning and signal processing, the Grassmannian has been applied to low-rank sparse matrix completion problems \cite{BOUMAL2015200}. One particularly interesting paper utilises a combination of SVD and Grassmann kernel learning for image classification through LDA, `tracing out' irrelevant data \cite{grassmanndiscriminant} by projecting. Grassmann learning is useful when optimisation on linear spaces only depends on subspace information, allowing one to lower complexity. In our strategy, it will play an important role as an auxiliary manifold, upon which gradient descent can be performed.  Much of this section is an elaboration and explanation of ideas from \cite{edelman1998geometry}.
\subsection{Definition}
The complex Grassmannian manifold $G(n,k)$ is the space of all $k$-dimensional subspaces of $\mathbb{C}^n$, endowed with a holomorphic structure. This generalises projective space by allowing all $k$-dimensional planes intersecting the origin. A point of $G(n,k)$ is specified by $k$ linearly independent vectors spanning the plane. Then we can give homogeneous coordinates by putting these into the columns of a (maximum rank) matrix. The plane is invariant under actions of $U(k)$, so we quotient our coordinates by the equivalence relation: 
\begin{equation}
W \sim \Tilde{W}, \; \textnormal{if and only if there exists} \; M \in GL(k,k) \; \textnormal{such that} \; W = \Tilde{W} M. 
\end{equation}
The invertible matrix $M$ is analogous to the scale factor in $\mathbb{P}^n$. 

The Grassmannian is a homogeneous space, since the unitary group acts transitively. Thus $G(n,k) \approxeq U(n)/(U(k)\times U(n-k))$, quotienting by the isotropy subgroup of the point spanned by the first $k$ standard basis elements. From this perspective, a point $[Q]$ looks like the equivalence class:
\begin{equation} [Q] = \bigg(
Q \begin{pmatrix}
Q_k & 0 \\
0 & Q_{n-k} 
\end{pmatrix}: Q_k \in U(k), \; Q_{n-k} \in U(n-k) \bigg).
\end{equation}
In contrast to the homogeneous coordinates approach, $Q$ will be an $n\times n$ unitary rather than an $n \times k$ matrix.

For comparison, consider the Stiefel manifold $V(n,k)$, the space of $k$-dimensional orthonormal frames in $\mathbb{C}^n$. Then $V(n,k) \approxeq U(n)/U(n-k)$, since to distinguish frames we no longer act with a $U(k)$ factor. A point $[Q]$ looks like:
\begin{equation} [Q] = \bigg(
Q \begin{pmatrix}
I_k & 0 \\
0 & Q_{n-k} 
\end{pmatrix}: Q_{n-k} \in U(n-k) \bigg).
\end{equation}
Unlike the Grassmannian, the Stiefel manifold can be intrinsically thought of as orthonormal $n \times k$ matrices, noting that $G(n,k) \approxeq V(n,k) / U(k)$. Our ultimate goal will be to compare subspaces of the $N_k$-dimensional complex vector space of global sections, $H^0(X, \mathcal{L}^k)$. To proceed, we must choose an appropriate metric on the Grassmannian. 

\subsection{The metric} \label{sec5.2}
To endow $G(n,k)$ with a metric, the homogeneous space perspective becomes useful.  This is because if the Riemannian geometry of a given manifold is well understood, the same is true for all quotients of the manifold. 

The tangent space to $G(n,k)$ at a point $[Q]$ can be identified with a subspace of $T_Q(U(n))$, using the notions of a vertical and horizontal spaces. The former contains vectors tangent to $[Q]$; the latter is chosen orthogonally, containing vectors of the form: \begin{equation} \label{ref22}
\phi = Q \begin{pmatrix}
Q_k & 0 \\
0 & Q_{n-k} 
\end{pmatrix}
\begin{pmatrix}
0 & -B\\
B^\dag & 0
\end{pmatrix}, \; \; Q_k \in U(k), \; Q_{n-k} \in U(n-k).
\end{equation}
Here, the third skew-symmetric matrix is a horizontal vector at the identity, shifted to the tangent space at $[Q]$ by premultiplication. In fact, this subspace is exactly what we wanted, and is isomorphic to $T_{[Q]}(G(n,k))$; intuitively, movement along the vertical space makes no difference to the quotient. Then the standard $U_n$ metric, restricted to \ref{ref22}, gives us a metric on the Grassmannian. Clearly it is totally determined at the origin.

Of course, for a given application, the metric chosen would ideally respect the geometry of the loss landscape. However, since we cannot a priori know what this is, we will use this (likely) suboptimal metric for computational efficiency. As a result, our metric choice will converge more slowly, but should still reach a global minimum (unless convexity is relaxed, which will unfortunately eventually happen). 
\subsection{Geodesics} \label{sec5.3}
Although we derived the metric using $U(n)$ equivalence classes, in practice we will work with $n \times k$ orthonormal representatives. A point of the Grassmannian now looks like:
\begin{equation}
[Y]=\{YQ_k, \; Q_k \in U_k\},
\end{equation}
using the fact that $G(n,k) \approxeq V(n,k) / U(k)$. The horizontal subspace to $[Y]$ contains $n \times k$ matrices $H$, such that $Y^\dag H=0$. Applying the previous metric gives the formula:
\begin{equation} \label{ref23}
Y(t) = (YV \; \; U) \begin{pmatrix}
\textnormal{cos} \; \Sigma t \\
\textnormal{sin} \; \Sigma t 
\end{pmatrix} V^\dag,
\end{equation}
for geodesics, where $Y(0)=Y$, $\dot{Y}(0)= H$, and  $U \Sigma V^\dag$ is the compact SVD of $H$. The matrix $Y$ is a Stiefel representative of the initial point and $H$ is the initial velocity vector. The cos and sin operations act elementwise along the diagonals, on the principal values. A derivation can be found in \cite{edelman1998geometry} (just substitute \ref{ref23} into the ODE defining geodesics). 

Importantly, using this formula allows us to explore the full Grassmann space whilst using Stiefel representatives. The geodesic equation will move around orthonormal representatives in a consistent way, using the horizontal directions. In the case of gradient descent, the choice of metric allows us to define descent directions by defining $\nabla$. Moreover, moving along geodesics by small $\epsilon$ in the extremal directions allows us to discretise our gradient paths, analogous to cutting up a curved path in $\mathbb{C}^n$ into short straight segments.

\section{Strategy} \label{ref14}
We return to K\"ahler-Einstein metrics. 

In sections \ref{sec1} through \ref{sec2}, we argued that naive applications of ML to K\"ahler manifolds can have many drawbacks, the most serious being that the tensors produced may fail to be metrics. In section \ref{sec3}, we saw that traditional methods, in particular Donaldson's algorithm, suffered seriously from the curse of dimensionality. As a result, we postulated a `no free lunch' theorem for numerical K\"ahler metrics. 

One would like combine these approaches to learn `geometrically'. We achieve this by embedding into a lower-dimensional $\mathbb{P}^{N_s}$ and approximating the potential there. This is is done in two different ways. Firstly, we use Donaldson's algorithm, learning balanced metrics for the subspaces. Secondly, we perform a joint `fibre bundle' optimisation on the product manifold: Stiefel $\times$ Symmetric positive definite matrices. The latter method means simultaneously optimising the basis and  corresponding $h$-matrix. 

In both cases, a suitable projection is identified with gradient descent, using the $\sigma$-error as a loss function. We find qualitative agreement between the behaviour of the resulting optimal solutions as the dimensionality of the subspace varies. Furthermore, we find that this method fits very well with the eigenfunction expansion interpretation of the potential.

Finally, we reiterate that our methods obviously rely on Monte Carlo integration. It is hard to avoid this in numerical geometry, but it is possible that it introduces inaccuracies in some cases, depending on the rate of convergence. In the CY case, we first sample with respect to the measure induced by the Fubini-Study volume form in projective space, then reweight by the ratio with $\textnormal{d}\mu_\Omega$. Therefore, the rate of convergence also depends on the $L^\infty$ norm of this $C^\infty$ reweighting function. However, this issue is the same for all existing methods discussed in this paper. 
\subsection{Projecting down}
Recall that the algebraic ansatz came from the Veronese embedding of $X$ with $\mathcal{L}^k$ into $\mathbb{P}^{N_k}$. Its high computational cost came from the rapid growth of the space of global sections, which scaled like $O(N_k^2)$.  A potential solution is to restrict to a subspace of $H^0(X, \mathcal{L}^k)$, via a projection from $\mathbb{P}^{N_k}$ to $\mathbb{P}^{N_s}$.  We thus define the map $I: X \rightarrow \mathbb{P}^{N_s}$, as:
\[
\begin{tikzcd}[row sep=2.5em]
 & \mathbb{P}^{N_k} \arrow{dr}{P} \\
X \arrow{ur}{E} \arrow{rr}{I} && \mathbb{P}^{N_s}.
\end{tikzcd}
\]
The projection can be specified by $N_s+1$ linear functions on $\mathbb{P}^{N_k}$, so that $I$ takes the form:

\begin{equation}
I: X \dashrightarrow \mathbb{P}^{N_s}, \; P({\textbf{z}}):=(T^0({\textbf{z}}):...:T^{N_s}({\textbf{z}})).
\end{equation}
Here, the $\{T^0$,..., $T^{N_s}\}$ are weighted linear combinations of degree $k$ monomials in the homogeneous coordinates. Provided they are linearly independent, these define a point of $G(N_k,N_s)$. We would like to use familiar machinery to produce K\"ahler metrics on $X$, via pullbacks of Fubini Study from $\mathbb{P}^{N_s}$. However, in order to do this, we must show that for a `random' choice $\{T^0,... ,T^{N_s}\}$, the map $I$ is an embedding. The proof is mathematically elementary, but for the sake of completeness and accessibility, we include it. 
\begin{theorem}{Let $X \subset \mathbb{P}^N$ be a nonsingular variety. The space of its embeddings is Zariski dense in the space of projections from $\mathbb{P}^N$ to $\mathbb{P}^{K}$, provided $K > 2n+1$.} \label{thm1}
\end{theorem}
\begin{proof}
Let $\Pi^{d}$ denote the spaces of point projections from $\mathbb{P}^d$ to $\mathbb{P}^{d-1}$, and let $\Pi^{N,K}$ denote the space of projections from $\mathbb{P}^N$ to $\mathbb{P}^{K}$. For the collection $\Pi^d$, Theorem \ref{thm1} is a standard result in algebraic geometry, by dimension counting \cite{Shafarevich:1596976}. Now consider the space of products: $\textnormal{Proj}:= \displaystyle \prod_{K<d\leq N} \Pi^d$. There is a surjective map $f$ (composition) from $\textnormal{Proj}$ to $\Pi^{N,K}$ (given an element of Proj defined by linear functions $E_0,...,E_K$, just extend to a full basis, then successively project out). Clearly the space of products of embeddings is dense in the Zariski topology on Proj. So its image under $f$ is dense in $\Pi^{N,K}$.
\end{proof}

\subsection{Balanced subspaces} \label{ref24}
Having selected a subspace [M], we consider its relation to Donaldson's algorithm. In particular, note that the notions \ref{ref6} and \ref{ref7} make sense for it alone. 

Firstly, and obviously, the $L^2$ inner product \ref{ref6} can be restricted to the set of sections $\{T^0$,.., $T^{N_s}\}$. Denote the span  by $\textbf{p} \subset H^0(X, \mathcal{L}^k)$. Given a bundle metric $h$, Hilb($h$) defines a Hermitian form on $\textbf{p}$. It is slightly harder to see that the other directions works. But it indeed holds, since given a Hermitian form $G$ on $\textbf{p}$, the pointwise condition \ref{ref7}  uniquely determines a line bundle metric, provided that $\{T^0$,..., $T^{N_s}\}$ is basepoint-free. In our case this will always be true. In other words, it makes sense to discuss a pair $(G,h_\infty(\textbf{p}))$, balanced with respect to $(\mathcal{L}^k, $\textbf{p}$)$. Note that again, $\textbf{p} \in G(N_k, N_s)$. Unfortunately we do not have a `subspace' result corresponding to Theorem \ref{ref20}. We have however numerically verified convergence, as will be seen in next section of this paper.
\subsection{Subspace potentials}
Consider the subspace Span$\{T^0,..., T^{N_s}\} \in G(N_k,N_s)$. This defines an algebraic potential:
\begin{equation} \label{ref12}
K_j([z])= \frac{1}{ \pi k} \textnormal{log} \; \frac{T^\alpha h_{a\bar{\beta}}\bar{T}^{\bar{\beta}}}{|z^j|^{2k}},
\end{equation}
provided we have chosen a positive Hermitian $h_{\alpha \bar{\beta}}\in P_{N_s}(\mathbb{C})$. For our first approach, the Grassmann-Donaldson algorithm, we select this matrix as the balanced metric corresponding to the subspace. This is well-defined by the discussion of the previous subsection. Now the connection to the Fourier mode viewpoint can be made clear. Extending $\{T^0,...,T^{N_s}\}$ to a full basis, we have, reproducing equation \ref{ref8}:
\begin{equation}
K - K_{FS} = \frac{1}{\pi k}\textnormal{log} \; (T^\alpha h_{\alpha \bar{\beta}} \bar{T}^{\bar{\beta}}) - \frac{1}{\pi} \textnormal{log} \; (\sum_i |z^i|^2) = \frac{1}{\pi k} \textnormal{log}  (\phi^{\alpha \bar{\beta}} h_{\alpha \bar{\beta}}),
\end{equation}
defining the new eigenfunctions $\phi^{\alpha \bar{\beta}}$ as linear combinations of the previous $\psi^{\alpha \bar{\beta}}$. Selecting a subspace of sections corresponds exactly to throwing away a collection of eigenfunctions $\phi^{\alpha \bar{\beta}}$ in the above expansion, ideally such that the corresponding coefficients are small. However, we would like to do this without training on the full space, which is computationally taxing. 
\subsection{Grassmann-Donaldson optimisation}

This can be summed up in the psuedocode of \textbf{Algorithm 1}.

\makeatletter
\def\BState{\State\hskip-\ALG@thistlm}
\makeatother

\begin{algorithm}
\caption{Grassmann-Donaldson}\label{euclid}
\begin{algorithmic}[1]
\State $\text{CY} \gets \text{Projective Calabi-Yau variety}$
\State $\mathcal{L}^k \gets \text{Positive line bundle with $N_k$-dimensional space of sections}$
\State $N_s \gets \text{Subspace dimension}$
\Function{$\mathcal{L}$}{\textbf{p}}: $G(N_k, N_s) \rightarrow \mathbb{R}$:
\State Compute h(\textbf{p}) by $12$ iterations of $T$ on a basis $\{T^0, ..., T^{N_s}\}$ for $\textbf{p}$
\State Compute potential $K(h(\textbf{p}))$, \Return $\sigma(K)$ by integration of K\"ahler form over CY
\EndFunction
\If {$N_k < N_s$} \algorithmicend
\Else 
\State    \text{Instantiate complex Grassmannian manifold $G(N_k,N_s)$}
\State \text{Choose $\textbf{p}_0 \in G(N_k, N_s)$} by random sampling
\State $k \leftarrow 0$
\While{STOP-CRIT \textbf{and} $k < k_{max}$}
\State $\textbf{p}_{k+1} \leftarrow \textbf{p}_k-\alpha_k \nabla \mathcal{L}(\textbf{p}_k)$ 
\State with $\alpha_k = \textnormal{min}_\alpha \mathcal{L}(\textbf{p}_k-\alpha \nabla \mathcal{L}(\textbf{p})) $
\State $k \leftarrow k+1$
\EndWhile
\State \Return $\textbf{p}_k$
\EndIf

\end{algorithmic}
\end{algorithm}

Essentially, having chosen a Calabi-Yau, a line bundle power, and a fixed subspace dimension, we identify a good restricted subspace metric without a priori knowledge, using relatively vanilla gradient descent on the Grassmannian. Our loss function $\mathcal{L}$ is just the sigma error corresponding to the algebraic potential, derived from the approximately balanced metric at a point of $G(N_k,N_s)$, in the sense discussed in Section \ref{ref24}. The K\"ahlericity of the metric for a `random' choice of $\mathbf{p} \in G(N_k,N_s)$ is guaranteed by the embedding result of Theorem \ref{thm1}.

For instantiation of the complex Grassmann manifold we used PYMANOPT \cite{pymanopt}, a Python package for optimisation on Riemmannian manifolds. We adapted this for a JAX framework. In the pseudocode, STOP-CRIT refers to a collection of stopping criteria, for example a minimum step size or gradient norm. The Riemannian gradient descent algorithm uses a line search method, noting that the $\nabla$ is defined by the choice of metric discussed in Section \ref{sec5.2}. As already discussed, this is likely suboptimal. For computation of the $T$-operator we utilised the Python package CYJAX \cite{cyjax}, adapting the code to allow the computation of the operator on a subspace of sections, encoded in a $N_s$ by $N_k$ orthonormal matrix representative. 

A serious problem is the necessity of passing the gradient through many nonlinear applications of the $T$-map. By default, the gradient will include point sampling and computation of the measure weights in the numerical evaluation of the loss. JAX allows us to manually take these steps out of $\nabla \mathcal{L}$. However, training on iterations of the $T$-operator is necessarily costly, reducing the computational advantage of the subspace approach. A related issue is that, as discussed in section \ref{sec3.6}, the Donaldson algorithm does not  optimise Ricci-flatness within a given algebraic class. It is therefore natural to consider other methods for jointly optimising for the subspace and $h$, other than the $T$-operator.
\subsection{Bundle (joint) optimisation}

We would like to optimise the basis of sections and $h$-matrix simultaneously. 

Recall that the tautological bundle $E$ over the Grassmannian, a subbundle of the trivial $G(N_k, N_s) \times \mathbb{C}^{N_k}$, is such that the fibre over a point $p$ contains the vectors in $\mathbb{C}^{N_k}$ belonging to it. We want to optimise over the space $\textnormal{Herm}(E)$ of Hermitian metrics on the fibres of $E$. However, this can be simplified by correctly choosing the right Stiefel representatives. Consider the following commutative diagram:

\[ \begin{tikzcd} \label{ref25}
f^*\textnormal{Herm}(E) \arrow{r}{h} \arrow[swap]{d}{\tilde{\pi}} & \textnormal{Herm}(E) \arrow{d}{\pi} \\%
V(N_k,N_s) \arrow{r}{f}& G(N_k,N_s)
\end{tikzcd}.
\]
It is useful to note that $f^*\textnormal{Herm}(E) \simeq V(N_k,N_s) \times P_{N_s}(\mathbb{C})$. Given a point of the pullback bundle $(p, H)$, where $H$ is Hermitian metric corresponding to the equivalence class $[p]$, evaluate $H$ on the frame to obtain a matrix (this is trivially an isomorphism).  We will use this to simplify the gradient exploration of Herm(E). 

We recap material covered earlier in the paper. Given a parameterised path in the Grassmannian, given by $\gamma(t): [0,1] \rightarrow G(N_k, N_s)$, there exists a unique lift to the Stiefel manifold $\tilde{\gamma}(t): [0,1] \rightarrow V(N_k, N_s)$, such that the following conditions all hold: $(i)$ $f\circ \tilde{\gamma}(t)=\gamma(t)$, for all $t$ $(ii)$ $\tilde{\gamma}(0)=p$ $(iii)$ $\tilde{\gamma}'(t)$ is horizontal, for all $t$. We have already seen this discussed in Sections \ref{sec5.2} and \ref{sec5.3}. Now we apply it to the situation of interest to us. Given a path now in the bundle $\textnormal{Herm}(E)$, there is a unique lift $(\tilde{\gamma}_1(t), \tilde{\gamma}_2(t))$ in the pullback $V(N_k, N_s) \times P_{N_s}(\mathbb{C})$, such that the first factor $\tilde{\gamma}_1'(t)$ is always horizontal, after fixing a starting point. So one can move in the fibre bundle $\textnormal{Herm}(E)$ by consistently choosing the right representatives. This is very easily performed in the PYMANOPT framework, because as discussed, points on $G(N_k, N_s)$ are already stored in frame matrices there.

\makeatletter
\def\BState{\State\hskip-\ALG@thistlm}
\makeatother

\begin{algorithm}
\caption{Bundle optimisation}\label{euclid1}
\begin{algorithmic}[1]
\State $\text{CY} \gets \text{Projective Calabi-Yau variety}$
\State $\mathcal{L}^k \gets \text{Positive line bundle with $N_k$-dimensional space of sections}$
\State $N_s \gets \text{Subspace dimension}$
\Function{$\mathcal{L}$}{\textbf{p}, h}: $\mathcal{L}: V(N_k, N_s) \times P_{N_s}(\mathbb{C}) \rightarrow \mathbb{R}$
\State Compute algebraic potential $K(\textbf{p}, h)$ for the frame-matrix pair $(\textbf{p}, h)$
\State \Return $\sigma(K)$ by integration of the corresponding K\"ahler form over CY
\EndFunction
\If {$N_k < N_s$} \algorithmicend
\Else 
\State    \text{Instantiate product manifold $V(N_k, N_s) \times P_{N_s}(\mathbb{C})$}
\State \text{Choose $\textbf{p}_0 \in V(N_k, N_s) \times P_{N_s}(\mathbb{C}$}) by random sampling
\State $k \leftarrow 0$
\While{STOP-CRIT \textbf{and} $k < k_{max}$}
\State $\textbf{p}_{k+1} \leftarrow \textbf{p}_k-\alpha_k \nabla \mathcal{L}(\textbf{p}_k)$ 
\State with $\alpha_k = \textnormal{min}_\alpha \mathcal{L}(\textbf{p}_k-\alpha \nabla \mathcal{L}(\textbf{p})) $
\State $k \leftarrow k+1$
\EndWhile
\State \Return $\textbf{p}_k$
\EndIf

\end{algorithmic}
\end{algorithm}

The bundle optimisation approach is outlined in the pseudocode of \textbf{Algorithm 2}. We perform gradient descent on the product manifold $V(N_k, N_s) \times P_{N_s}(\mathbb{C})$, using the standard product metric \cite{chavel1984eigenvalues}, whilst moving in horizontal directions infinitesimally. For clarification, the algebraic potential corresponding to a frame matrix pair $(\textbf{p},h)$ is just \ref{ref1}, with monomials and coefficients corresponding to $\textbf{p}$ and $h$ respectively. The goal is to identify the point $(\textbf{p}_{opt}, h_{opt})$ such that the corresponding metric has minimal error. 

For $N_s=N_k$, we recover gradient descent on the full $h$-matrix alone, an approach which is very similar to previous work in \cite{headrick} and \cite{moduli}. However, in this case our method searches exclusively on the manifold of positive definite Hermitian matrices. Our approach also makes choosing a new metric straightforward. Because there is no need to pass the gradient through the $T$-operator, the joint optimisation approach is dramatically quicker than Grassmann-Donaldson optimisation. By directly optimising for Ricci-flatness, $\sigma$-errors of lower orders of magnitude become achievable. We discuss the results of both methods in the next section.

\section{Results}
For the numerical experiments of this paper, we will focus on the Dwork threefolds of equation \ref{ref9}. The algebraic metrics and Donaldson approach can easily be extended to arbitrary polynomial hypersurfaces and other generalisations (CICYs and quotients) by computing an appropriate basis of global sections \cite{Braun}. For $n=5$ and  $\phi=0$, it is the Fermat quintic, upon which Donaldson's methods have already been tested on in some depth \cite{Braun_2008}, \cite{moduli}, \cite{Braun}. The singularities occur for $\phi$ equal to five times a fifth root of unity. 

We chose random initial points on the Grassmann manifold by choosing the frame representative as i.i.d standard normal. This gives a uniform distribution with respect to the measure induced by our chosen metric \cite{chikuse2012statistics}. For the joint approach, we experimented with several distributions on the space of Hermitian positive definites, for example the complex Wishart, or a QR decomposition of a random Gaussian. For each evaluation of the loss function, a Monte Carlo integral, we sampled a new set of at least $30,000$ points to prevent overfitting. We then tested the resulting optimal metrics on a new set of at least $30,000$ sampled points (in both cases, typically more).

\begin{figure}
\subfloat[\centering O(4) with linear scale.]{%
  \begin{minipage}{0.49\textwidth}
    \includegraphics[width=\linewidth]{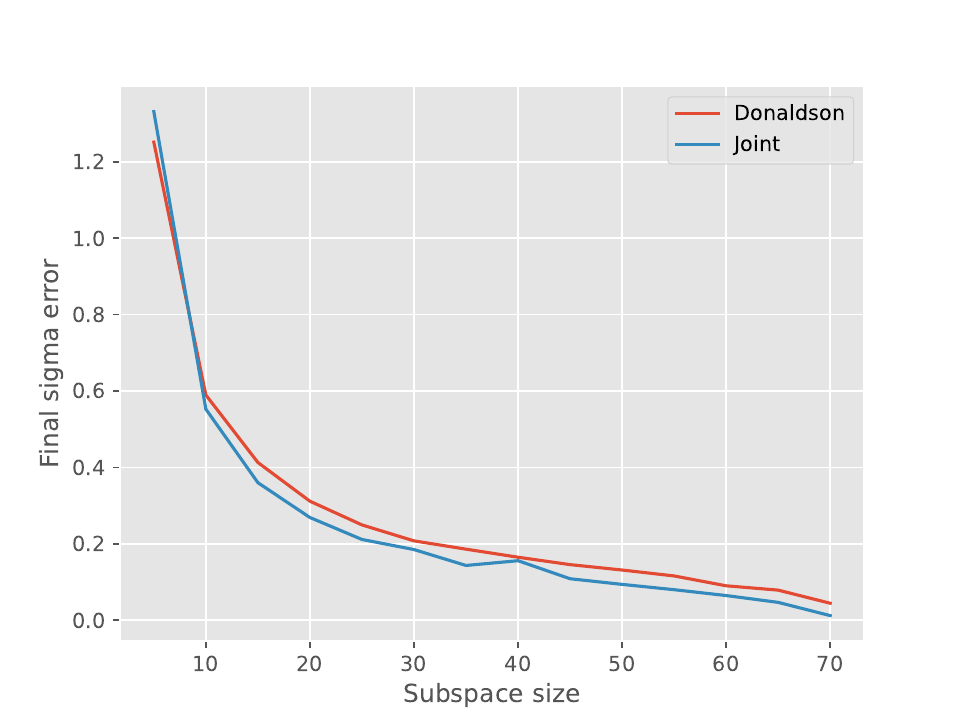} \label{fig6}
  \end{minipage}}
 \hspace{1mm}
  \subfloat[\centering O(5) with linear scale.]{%
  \begin{minipage}{0.49\textwidth}
    \includegraphics[width=\linewidth]{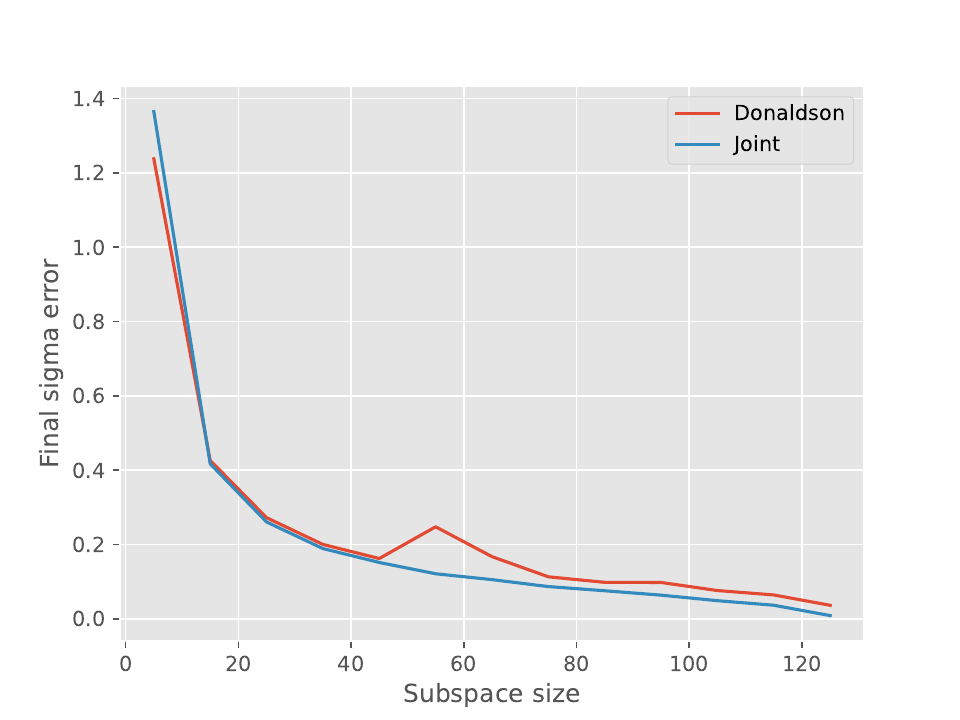} \label{fig3}
  \end{minipage}}
 \\
  \subfloat[\centering O(6) with linear scale.]{%
  \begin{minipage}{0.49\textwidth}
    \includegraphics[width=\linewidth]{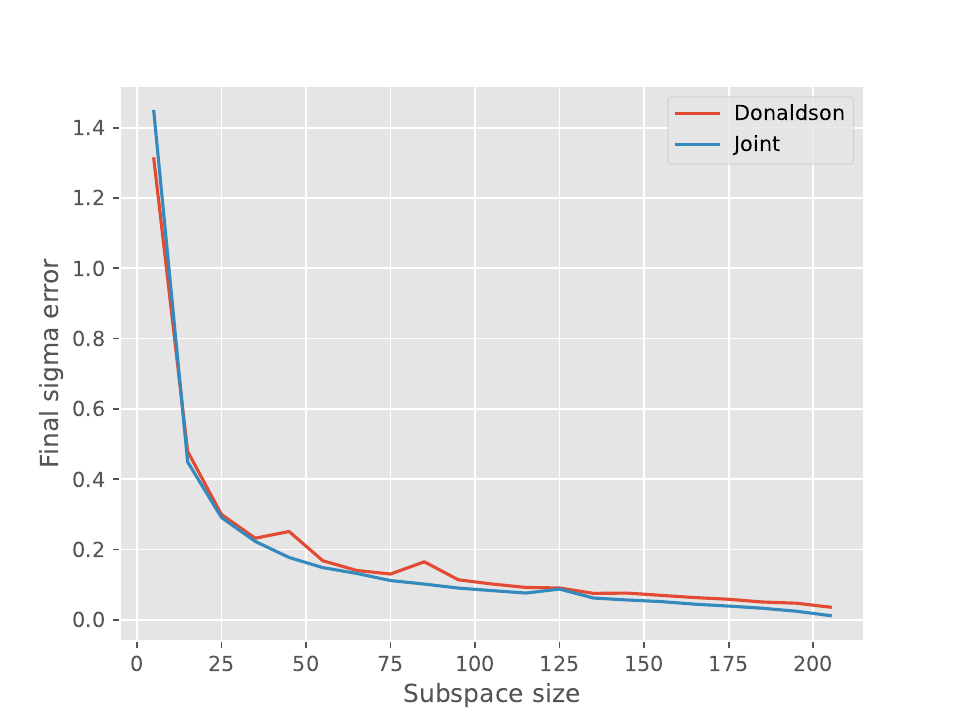} \label{fig4}
  \end{minipage}}
 \hspace{1mm}
  \subfloat[\centering O(4) with log scale.]{%
  \begin{minipage}{0.49\textwidth}
    \includegraphics[width=\linewidth]{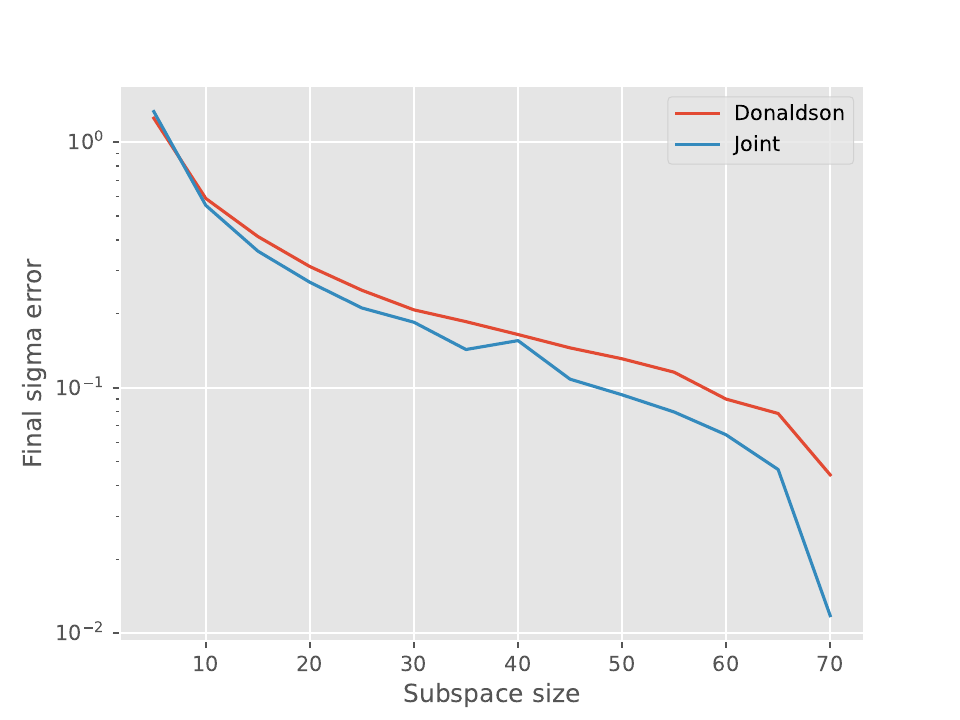} \label{fig5}
  \end{minipage}}
  \caption{Final test $\sigma$-errors for Grassmann-Donaldson and bundle optimisation on the Fermat quintic and a range of line bundles.} \label{fig7}
  \centering
  \end{figure}

\subsection{The Fermat quintic}
We start with the simple case of the quintic. The results are shown in Figure \ref{fig7}, for the line bundles $\mathcal{O}(4)$, $\mathcal{O}(5)$, and $\mathcal{O}(6)$. Note that the majority of plots in this paper have subspace dimension on the $x$-axis, so that \emph{plotted points correspond to optimised errors}. 

It is clear that a great deal of learning can happen on the subspaces, since the majority of the lost error occurs within a small fraction of sections. For example, working with the $\mathcal{O}(6)$ line bundle, an $O(10^{-2})$ $\sigma$-error is achieved with just one half of the total dimension. Then there is a shallow, roughly linear decrease for the remaining growth up to the full $N_k$. Considering the significant scaling of computational cost with the growth of $N_s$, this is a positive result, justifying the use of subspace methods. One can imagine choosing a subspace and bundle dimension tailored to a desired degree of metric accuracy within this framework. 

There was no reason to expect the plots to take these concave forms. Many behaviors could have occurred as the basis size was varied from $0$ to $N_k$, for example a linear decrease, or no significant significant improvement before the full dimension. In our view, the concave behaviour is mathematically interesting and tells us something nontrivial about the embedding method. Naturally, the joint bundle optimisation outperforms the Grassmann-Donaldson method for all values of $N_s$, since the space of matrices in the former case completely contains the latter. The only thing that could obstruct this is the loss landscape, where local minima could prevent convergence to the desired solution. 

\begin{figure}
\subfloat[]{%
  \begin{minipage}{0.49\textwidth} 
    \includegraphics[width=\linewidth]{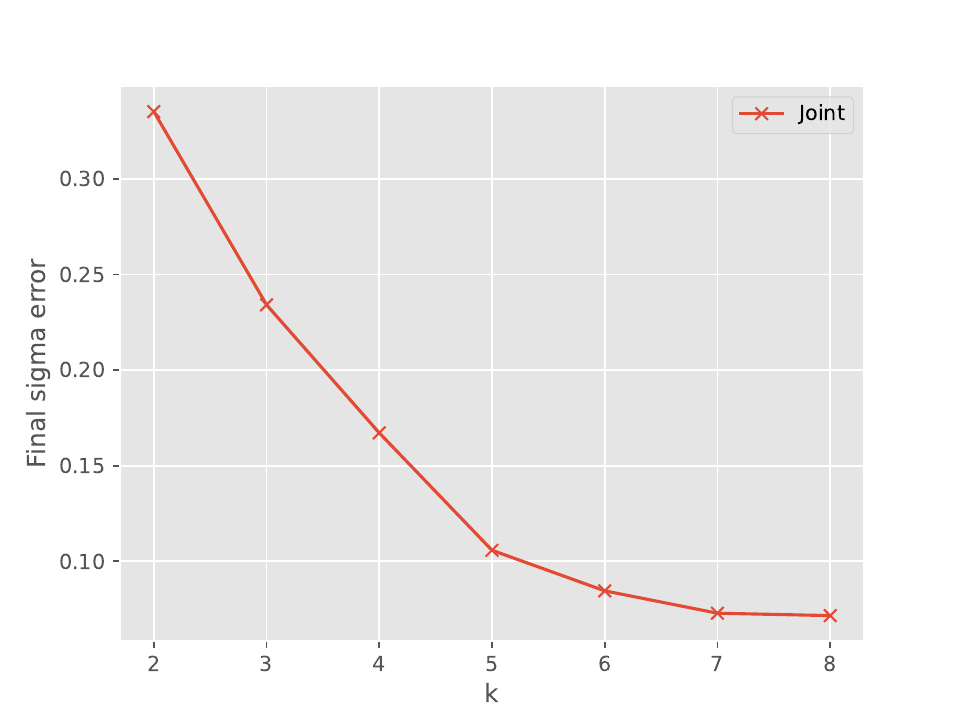} \label{fig1}
  \end{minipage}}
 \hspace{1mm}
  \subfloat[]{%
  \begin{minipage}{0.49\textwidth}
    \includegraphics[width=\linewidth]{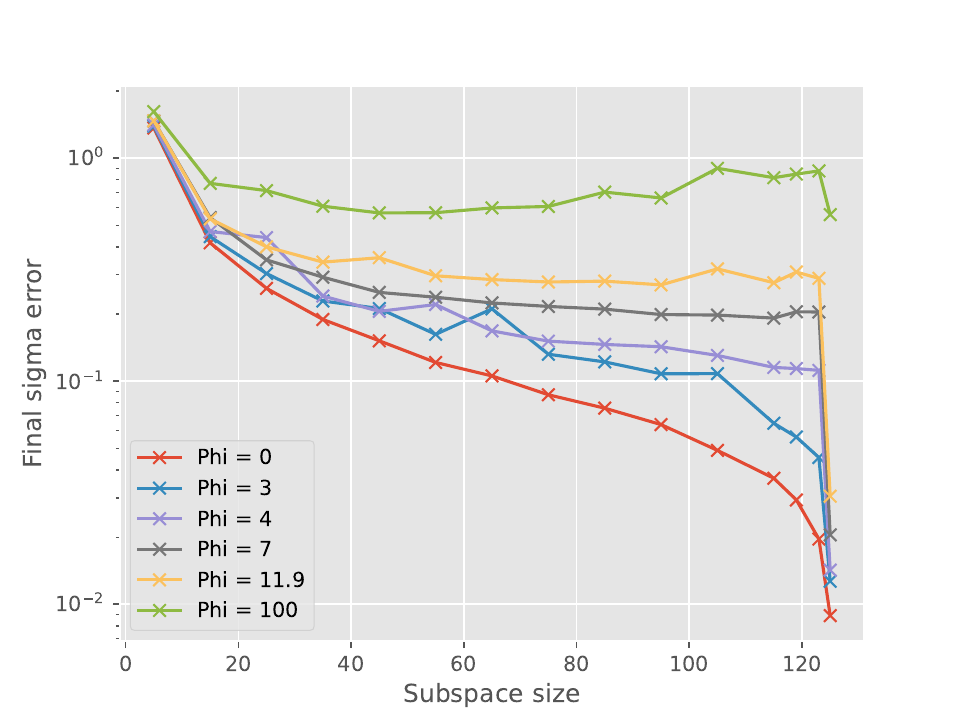} \label{fig2}
  \end{minipage}}
  \centering
\caption{$(a)$ Final test error for bundle optimisation and a fixed fraction $N_s=\frac{1}{2}N_k$, on the Fermat quintic. The $y$-axis has a linear scale. $(b)$ Final test errors using bundle optimisation for $\mathcal{O}(5)$ and a range of real moduli parameters on the Dwork. The initial positive matrix is chosen by a QR decomposition and the $y$-axis has a logarthmic scale.}
  \end{figure}

Another finding is that the loss curves for both approaches have essentially the same overall shape. The qualitative agreement between these two approaches confirms the existence of a lower-dimensional geometric structure in the space of global sections. If the $\sigma$-error is plotted in log scale, one observes the same pattern for all bundles, that the relative performance difference between the approaches grows as $N_s$ approaches $N_k$.

Note also that the loss curves are pushed into the origin as the tensor power of the bundle increases. More precisely, for a fixed fraction of global sections, i.e. $N_s = \gamma N_k$, where $\gamma \in [0,1]$, the $\sigma$-error seems to decay with $k$. We confirmed this in Figure \ref{fig1}, using a joint approach for half the total dimension as $k$ varies from $2$ to $8$. The `squeezing' behaviour suggests the possibility of a sequence of fixed fractional subspaces $S_k$, such that the corresponding restricted metrics still converge to Ricci-flatness in the limit. 
\subsection{Non-zero moduli parameters}
We consider the more general case. Figure \ref{fig2} shows the results of bundle joint optimisation on the Dwork for a range of moduli parameters. Note the apparent emergence of local minima in the manifold $V(N_k, N_s) \times P_{N_s}(\mathbb{C})$ with the growth of $\phi$. For example, for $\phi=11.9$, this is suggested by the relatively stable performance in the range $30 \leq N_s < 125$, with a sudden drop in error upon reaching the full space of sections. By $\phi=100$ highly variable outcomes are achieved for all dimensions. 

\begin{figure}
\subfloat[]{%
  \begin{minipage}{0.49\textwidth} 
    \includegraphics[width=\linewidth]{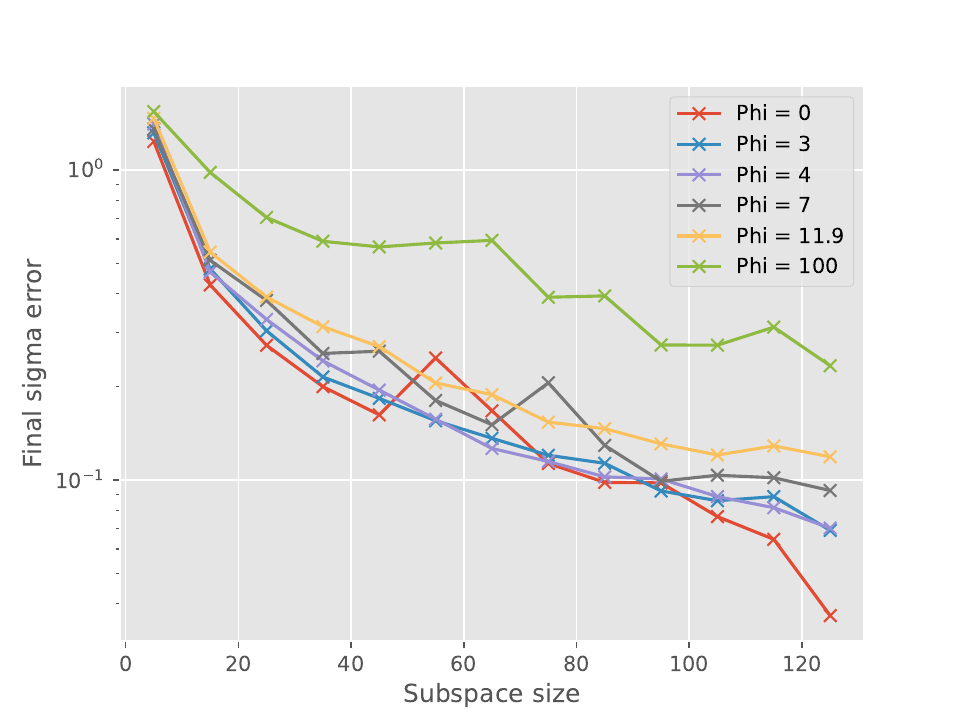} \label{fig9}
  \end{minipage}}
 \hspace{1mm}
  \subfloat[]{%
  \begin{minipage}{0.49\textwidth}
    \includegraphics[width=\linewidth]{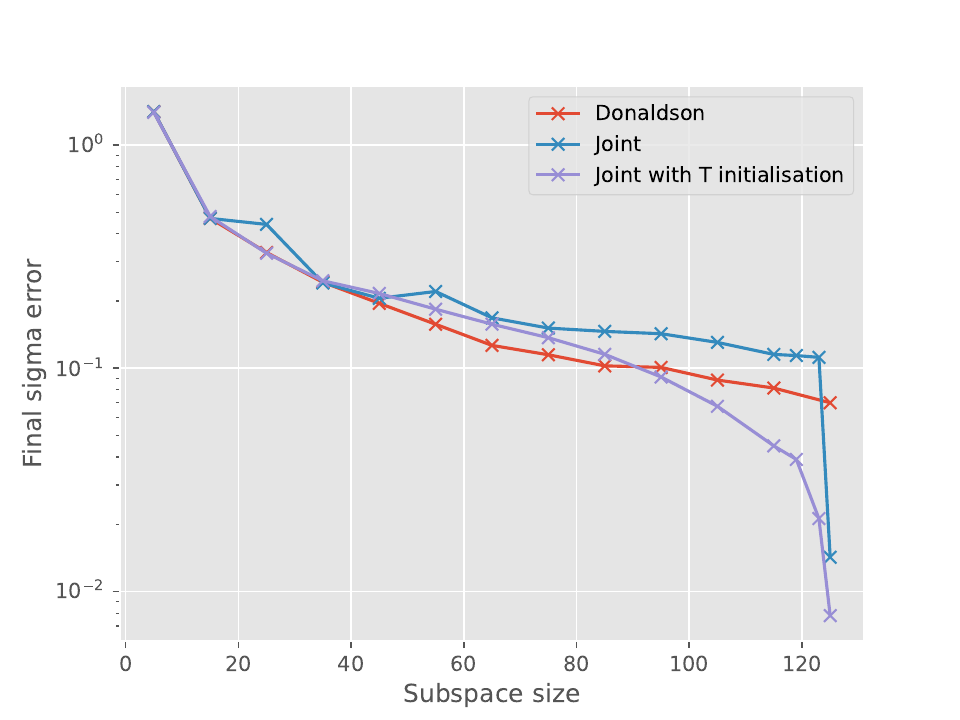} \label{fig10}
  \end{minipage}}
  \centering
  \\
    \subfloat[]{%
  \begin{minipage}{0.49\textwidth}
    \includegraphics[width=\linewidth]{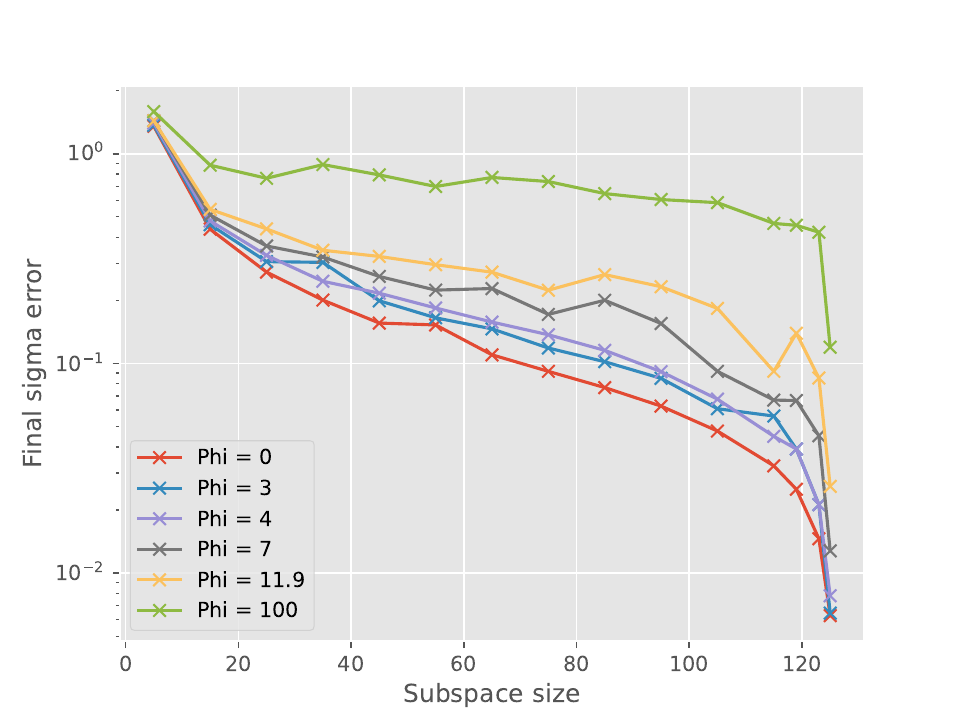} \label{fig11}
  \end{minipage}}
  \centering
\caption{All plots are for the Dwork equipped with the $\mathcal{O}(5)$ bundle, with a logarithmic scale on the $y$-axis. (a) Final test error for Grassmann-Donaldson optimisation for a range of real moduli parameters, initialising with QR distribution. $(b)$ Comparison of Grassmann-Donaldson, joint, and $T$-initialised joint optimisations for the $\phi=4$ case. $(c)$ Final test error for joint optimisation with a $T$-initialisation, for a range of real moduli parameters.}
  \end{figure}

The local minima did not occur for the Grassmann-Donaldson approach, as shown in Figure \ref{fig9}. Whilst the algorithm performs worse with the growth of $\phi$, there is still a continued decrease in optimised error, without plateauing. The implication is that the local minima are in the product manifold, rather than the Grassmannian. As shown in Figure \ref{fig10}, the Grassmann-Donaldson algorithm outperforms the joint approach on random initialisations for the $\phi=4$ case, supporting this intuition, since the space of solutions in the latter contains the former. Similar results were obtained for all sufficiently large $\phi$. 

To test this, we iterated the $T$-operator once on the initially sampled $h$-matrix, hoping to escape basins of attraction of the local  minima. The results for this are shown in Figure \ref{fig11}. Joint optimisation now improves with the growth of the subspace (no plateauing). It outperforms the Grassmann-Donaldson approach for a sufficiently large subspace dimension, and the randomly initialised joint approach in all cases (this is for $\phi=4$ in \ref{fig10}, but the same pattern occurred for any sufficiently large $\phi$). So a single initial $T$ step helped to avoid the local minima, without entirely solving the problem. This is not surprising. When the gradient descent moves to a new frame, the $h$-matrix T-initialisation corresponding to the first basis may no longer be a `good' choice.

The performance of Donaldson's algorithm with the growth of the complex structure parameter was discussed in \cite{cyjax}. We note that there, the final performance is essentially circularly symmetric in the naive distance from the origin, which makes intuitive sense. However, noting that the natural notion of distance on this space, the Weil-Petersson metric, is not circularly symmetric (see Figure 8 in \cite{CANDELAS199121}, there is only a five-fold rotational symmetry), we find this surprising, although it may be explainable by a lack of resolution. A comparison of its performance on a collection of five random quintics was presented in \cite{Braun}. However, to our knowledge a wider comparison of different methods on a variety of CYs is currently lacking in the literature. There may be some connection to the symmetry group of the CY, also acting on the space of global sections. 
\subsection{Summary}
We have presented implementations of the Grassmann-Donaldson and bundle approaches, for the Dwork family with a range of real moduli parameters \footnote{$\textnormal{Im}(\phi) \neq 0$ did not change the qualitative behaviour.}. The key finding is that for the $\mathcal{O}(k)$ bundles, a relatively small fraction of sections performed well in terms of the optimised error. We note that the errors are not directly competitive with the state-of-art neural networks. However, our approach is K\"ahler by construction, offering a solution to the problem arising from the failure of machine-learned metrics to enforce positivity. Moreover, we feel that the method is of some geometric interest in its own right. By using a straightforward gradient descent approach, we were able to identify a good subspaces of sections without a priori knowledge. Moreover, we observed continued improvement as $k$ increases for all fixed fractions $N_s/N_k$. Finally, as the parameter $\phi$ was varied, we observed the occurrence of local minima in the loss landscape of the manifold $V(N_k, N_s) \times P_{N_s}(\mathbb{C})$. This problem was partially mitigated by initialising with a $T$ step.

\section{Discussion} \label{ref15}
In the introduction, we discussed the history of the K\"ahler metrics problem, highlighting how physical calculations, such as the work of Candelas and collaborators \cite{CANDELAS199121}, have guided mathematical intuition in this area. Remarkably, Erich K\"ahler already knew in 1933 that the Ricci curvature of his proposed metrics was locally expressible in terms of a real-valued function \cite{Kahler}. He was also aware that the cohomology of the corresponding forms was a topological invariant of the manifold. As noted by the geometer Jean-Pierre Bourguignon, `more or less every page, he (K\"ahler) opens a new path that has later turned out to be crucial for the development of the subject' \cite{slides}. As he defined his manifolds, K\"ahler was already considering them as solutions to Einstein's equations, since the simplicity of the Ricci curvature implies a reduced form for the vacuum EFE. This is echoed in Yau's own testimony. After struggling to understand the Ricci curvature in general relativity, he came to K\"ahler manifolds initially for the intuition they provided on this tensor \cite{hosono}. Thus from its beginnings K\"ahler geometry sits at the intersection of mathematics and theoretical physics.  

The original motivation for numerical Ricci-flat metrics is string phenomenology. Speaking broadly, its long-term goal is the identification of higher-dimensional geometries which recover four-dimensional theories, agreeing with observed particle physics in their low energy limits. Efforts in this direction originally proceeded using simple geometries for which the metric was known, or sophisticated models for which it wasn't \cite{Candelas}, \cite{Douglas_2015}, \cite{hermitian}. In the former case, the simplicity of the compactifying spaces meant that they struggled to reproduce the complexity of the standard model. In the latter, various tricks became necessary to avoid the metric. Due to recent ML progress, contact has been made between the geometric and computational sides of string phenomenology. We see our work within this tradition, learning geometrically in a concrete way. 

We include some further remarks on phenomenology, relevant for future research avenues. Beyond the gauge group $SU(3) \times SU(2) \times U(1)$ \footnote{For mathematicians and computer scientists, these roughly correspond to the strong, weak, and electromagnetic forces.}, other observables can be deduced along mostly geometric lines, within a string-theoretic framework. Reproducing the correct number of particle generations \footnote{This is a categorisation of fermions into three groups with the same overall irreducible representations \cite{woit2017quantum}.} can impose various conditions on the pair $(M, \mathcal{V})$. ML can also contribute in this direction, for example by showing that approximately correct flavour hierarchies are reproduced away from symmetric points in moduli space \cite{precisionstringphenomenology}. The methods used here can be generalised to CICYs and quotients by computing the appropriate bases of sections. Donaldson's algorithm has already been applied to Hermitian Yang Mills connections through the `generalised' T-operator \cite{hermitian}. We believe that the Grassmann-Donaldson approach could also be used there, although the computational times may not be feasible. We are in the (very) nascent stages of `precision string phenomenology', although as of now, it is still  a `needle in a haystack search' \cite{quanta}.

It would be reductive to view such methods as exclusively applicable in string theory. The problem of numerical K\"ahler-Einstein manifolds is an interesting testcase for the application of machine learning in numerical geometry, PDEs, and theoretical physics. Balanced metrics have been extensively studied within a purely mathematical context, a major reason being their links to algebo-geometric notions of stability. In particular, under appropriate conditions on a polarized projective manifold $(X, \mathcal{L})$, the existence of a balanced metric for some power $\mathcal{L}^k$ is equivalent to Gieseker stability \cite{wang1}, \cite{wang2}. This becomes important when constructing `good' moduli spaces of vector bundles over projective manifolds, in a suitable sense. The asymptotic expansion of $\rho$ has been physically motivated with the path integral. It has also been applied to the study of BPS black holes compactified on a CY manifold \cite{blackholesbalancedmetrics}, \cite{integralkernel}. Intriguingly, an exact link is drawn there between maximal entropy and balancedness. It is our hope that the numerical methods developed here could be used in these contexts. 

The computation of Ricci-flat metrics can also be thought of as a geometric problem alone. We believe that there is a relationship between the following: $(i)$ the existence of a subspace structure on $H^0(M, \mathcal{L}^k)$ approximating Ricci-flatness well and $(ii)$ the underlying geometry, i.e. as a point in a parameterised moduli space. Our results suggest a correlation between the size of the symmetry group of  $M$ and the existence of such a lower-dimensional structure. There may be further work to do in untangling the relationship between balancedness and Ricci-flatness. In Donaldson's papers \cite{Donaldson}, we feel that this emerges purely in the proof, without clear geometric or physical intuition. There may be direct links from the geometry back to physics, since the data of a positive $\mathcal{L}$ over $M$ gives a geometric quantization of the Poisson structure corresponding to $\omega$. 

As previously mentioned, the asymptotic expansion of the density of states function $\rho(\omega)$ could provide a convergence guarantee for the Grassmann-Donaldson method. For example, there may exist a sequence of `fractional' expansions, converging to constant scalar curvature in an appropriate limit. An improvement of our algorithm would be an adaptive approach, whereby a numerical K\"ahler metric computed after a particular step of gradient descent is used to redefine a more relevant metric on the loss landscape. Finally, the embedding method necessarily depends on the existence of complex and K\"ahler structures. One may hope that the ideas of Donaldson, Yau, and Tian could be applied to real, Lorentzian manifolds in numerical relativity. However, there are many reasons to doubt this, since the K\"ahler condition implies remarkable simplifications on the geometry. In the words of Eugenio Calabi, he studied K\"ahler manifolds for one main reason: `because they are so simple' \cite{slides}. 

\section{Acknowledgments}

We thank Daattavya Aggarwal, Per Berglund, Mario Garcia-Fernandez,  Tristan H\"ubsch, Vishnu Jejjala, Damian Kaloni Mayorga Pe\~na, Minhyong Kim, Neil Lawrence, Justin Tan, and in particular, Jack Smith, for helpful conversations. We are also grateful to the anonymous referees for their careful reading of the text, and their helpful comments. Finally, we thank Aditya Ravuri and Haoting Zhang for assistance with programming and use of the Cambridge HPC. Oisin Kim and Challenger Mishra are supported by the Accelerate Program for Scientific Discovery.

\appendix
\section{Proof outline to Theorem \ref{ref19}}
\label{ref21}
The basic idea is to use a density of states functional:
\begin{equation} \label{ref36}
\rho(\omega) = \sum_i ||s_i||^2,
\end{equation}
where the sum is taken over a basis of orthonormal sections, with respect to the norm from $\omega$. This function is constant exactly when $\omega$ is balanced, and has an asymptotic expansion in $k$ involving the scalar curvature. Assuming it, arguments showing that a sequence of balanced $\omega_k$ converge to the Ricci-flat metric are standard \cite{Scalar}, \cite{hermitian}; we recap them later.

Let us consider the expansion itself. We note that existing analytic arguments have been presented in \cite{zelditch}, \cite{Catlin1999}, \cite{lu}; however, it is interesting to see how ideas from local index theory may be used to derive an alternative proof. We consider this now, in the hopes that it will provide its own insight. The theorems and results referenced in this Appendix without proof can all be checked in \cite{gilkey1984invariance}.

 Denote the space of $E$-valued $(0,k)$-forms $\mathcal{C}^\infty(\Lambda^{0,k}M \otimes E)$ by $\Omega(E)$. There then exists the usual Dolbeault complex:
\begin{equation} \label{ref28}
\Omega^0(M)\xrightarrow{\bar{\partial}} \Omega^1(M) \xrightarrow{\bar{\partial}}... \xrightarrow{\bar{\partial}} \Omega^d(M),
\end{equation}
with its associated cohomology, denoted by $H^q(M,E)$. 

Now, \ref{ref28} can be viewed as the simpler complex, with a corresponding index:
\begin{equation} \label{ref29}
\Omega^{even}(E) \xrightarrow{D} \Omega^{odd}(E), \end{equation} \begin{equation} \textnormal{Index}(D):= \textnormal{dim}(\textnormal{Ker}D) -\textnormal{dim}(\textnormal{Coker}D),
\end{equation}
where we have defined $D:=\bar{\partial}+\bar{\partial}^*$. This index
is equal to the alternating sum:
\begin{equation}
\Xi(M, E) := \sum_{k=0}^m(-1)^k \;\textnormal{dim}(H^k(M,E)),
\end{equation}
known as the holomorphic Euler characteristic. A famous result is:
\begin{theorem}{(Hirzebruch-Riemann-Roch).} \label{ref31} We have that:
\begin{equation}
\Xi(M,E) =\int_M \textnormal{td}(M)\textnormal{ch}(E), \end{equation}
where \textnormal{td}(M) is the Todd class of $TM$ and \textnormal{ch} is the Chern character of $E$.
\end{theorem} Although we will not precisely define them, we have the concrete expansions:
\begin{equation}
\textnormal{td}(M) = 1+\frac{c_1(M)}{2}+\frac{c_1^2(M)+c_2(M)}{12}+.... \;,
\end{equation}
\begin{equation}
\textnormal{ch}(E) = 1+c_1(E)+\frac{c_1(E)^2-2c_2(E)}{2}+... \;.
\end{equation}
where $c_i(E)$ and $c_j(M)$ are the Chern classes. 

Now, associated to \ref{ref29} are the even and odd heat kernels:
\begin{multline}
    \label{ref32}
K_t^+(x,y):=\sum_i e^{-t\lambda^{+}_i} \big( \sum_j s_{i,j}^+(x)\otimes (s_{i,j}^+)^*(y) \big), \\ K_t^-(x,y):=\sum_i e^{-t\lambda^{-}_i} \big( \sum_j s_{i,j}^-(x)\otimes (s_{i,j}^-)^*(y) \big),
\end{multline}
where the collections $s^{\pm}_{i,j}$ over $j$ are bases of $E$-valued forms corresponding to the $\lambda^{\pm}_i$ eigenspaces of the Laplacians $\Delta+:=D^*D$ and $\Delta-:=DD^*$. The sections are chosen to be orthonormal with respect to a fixed K\"ahler form $\omega$, whilst for notational cleanness, we suppress a second sum over them.

We are interested in a particular part of $K_t^+(x,y)$. Specialise to the case that $E=\mathcal{L}^k$, using the K\"ahler form $\omega$ induced by the Hermitian $\mathcal{L}$, and consider the $\lambda_0=0$ eigenspace only. This gives:
\begin{equation} \label{ref30}
B_k(x,y):= \sum_{j=0}^{N_k-1} s_j^{k+}(x) \otimes (s_j^{k+})^*(y)
\end{equation}
where $N_k=$ dim $H^0(M,\mathcal{L}^k)$, and is known as the Bergman kernel \cite{Catlin1999}, and we have included a $k$ index on the sections to make the bundle dependence explicit. Note that a priori there could also be contributions from $H^2(M, \mathcal{L}^k), H^4(M, \mathcal{L}^k)$... , by the Hodge Theorem for an elliptic complex. However, in this case, these all vanish for positive $\mathcal{L}$ and Calabi-Yau $M$, by the Kodaira vanishing theorem (because $K_M$ is trivial). To see the connection to our discussion, we take the trace of \ref{ref30}, yielding exactly \ref{ref36}, the density of states functional.

We still want an asymptotic expansion. What this means is that we allow the bundle $\mathcal{L}^k$ to vary, taking $k \rightarrow \infty$, whilst keep $\omega$ fixed. Let us take the traces of $K^{\pm}_t(x,x)$, denoting these by $K^{\pm}_t(x)$. Then these have their own asymptotic expansions in $t$:
\begin{equation} \label{ref26}
K^+_t(x) \sim a^+_{-d}(k)t^{-d}+a_{-d+1}^+(k)t^{-d+1}+... \; , \; \; \; \; K^-_t(x) \sim a^-_{-d}(k)t^{-d}+a_{-d+1}^-(k)t^{-d+1}+... \;,
\end{equation}
which we will use shortly. The notation of $k$-dependence for the functions $a$ emphasises the dependence on the bundle $\mathcal{L}^k$, which will vary. Since we want an asymptotic expansion in $k$, at this point we will make the substitution $t=\frac{1}{k}$ into \ref{ref26}.

Importantly, Theorem \ref{ref31} has a local analogy.
\begin{theorem}{(Local Hirzebruch-Riemann-Roch).} \label{ref27} We have that:\begin{equation}\lim_{t\rightarrow 0} \; (K^+_t(x)-K^-_t(x)) \;  \textnormal{dvol}_\omega=[\textnormal{Ch}(\mathcal{L})\textnormal{Td}(X)]_{2d},\end{equation}
where $[.]_{2d}$ denotes the $2d$-form term in the expansion.
\end{theorem}
Note that integration over $M$ recovers Theorem \ref{ref31}. It was shown in terms in a famous paper of Patodi \cite{Atiyah1973} that the negative power terms in \ref{ref26} all cancel in the LHS of Theorem \ref{ref27}; this has commonly been referred to as a `miraculous' calculation. Thus the LHS should be interpreted as the constant term in the expansion \ref{ref32} (recall this is just $H^0(M, \mathcal{L}^k)$, plus any contributions from higher order eigenspaces which may contribute in the limit $t\rightarrow 0$, i.e. $k\rightarrow \infty$. This can be expressed as:
\begin{equation}
\sum_{j=0}^{N_k-1}||s_j^{k+}(x)||^2+C(k),
\end{equation}
where $C(k)$ here denotes $k$ dependence, without telling us anything about asymptotic behaviour yet. 

Specialising to threefolds, Theorem \ref{ref27} implies that:
\begin{multline}
(\rho_k(\omega)+C(k)) \; \textnormal{dvol}_{\omega} = \frac{k^3(c_1({\mathcal{L}}))^3}{3!}+\frac{k^2(c_1(\mathcal{L}))^2\wedge c_1(M)}{4}+O(k) \\ = \frac{k^3\omega^3}{3!(2\pi)^3}+\frac{k^2\omega^2\wedge \rho}{2(2\pi)^3}+O(k)= \frac{\textnormal{dvol}_\omega}{(2\pi)^3}\bigl(k^3+\frac{1}{2\pi}S(\omega)+O(k) \bigr),
\end{multline}
exactly what we wanted to show, that is:
\begin{equation} \label{ref33}
\rho_k(\omega)=\sum_{j=0}^{N_k-1} ||s^+_j(x)||^2 \sim \frac{1}{(2\pi)^d}\bigl( k^d + \frac{1}{2\pi}S(\omega)k^{d-1}+O(k^{d-2}) \bigr),
\end{equation}
provided that the $C(k)$ terms only contribute at $O(k)$ - in this paper we will take this as a given. To verify this analysis for the general $d$ case one would require using higher order Chern characters, which may be difficult.

So now, we can use this expansion to show that a sequence of balanced metrics on tensor power bundles - $(\omega_k, \mathcal{L}^k)$ - converges to the Ricci-flat metric. The asymptotic expansion \ref{ref33} made sense for any $\omega$ coming from $\mathcal{L}$. We specify to balanced $\omega_k$, so that $\rho_k(\omega_k)$ is constant. The value of this constant is determined by integration: since the $s_i(.)$ are an orthonormal basis we must have that $\int_X \rho_k(\omega)=N_k$ for any $\omega$, so that for the balanced metrics $\rho_k(\omega_k)=\frac{N_k}{V}$, where $V$ is the CY volume. Moreover, we have the usual Riemann-Roch expansion for $N_k$:
\begin{equation} \label{ref34}
N_k = a_0k^d +a_1k^{d-1}+...=\textnormal{Vol}(X)\cdot k^d+\frac{1}{2\pi}\int_X S(\omega) \cdot k^{d-1}+... \;.
\end{equation}

The definition of the `asymptotic' expansion (see \cite{Scalar}, \cite{Scalarii}) means that:
\begin{equation}
\big\Vert \rho_k(\omega_k) - k^d -\frac{1}{2\pi}S(\omega_k)k^{d-1}\big\Vert_{C^0(X)} \leq C k^{d-2},
\end{equation}
for some constant $C$, which has absorbed the $\frac{1}{(2\pi)^d}$ prefactor. Moreover, we know that $\rho_k(\omega_k)$ has constant value $\frac{N_k}{V}$, and we can expand the numerator using \ref{ref34}, yielding:
\begin{equation}
\big\Vert \frac{1}{V}(Vk^d+a_1k^{d-1}+...)-k^d-\frac{S(\omega_k)}{2\pi}k^{d-1} \big\Vert_{C^0(X)} \leq Ck^{d-2}
\end{equation}
which means that:
\begin{equation}
\big\Vert \frac{2\pi}{V}a_1-S(\omega_k) \big\Vert_{C^0(X)}=O(k^{-1}),
\end{equation}
implying that the curvature tends to a constant value in the limit. On a CY manifold this implies zero scalar curvature and that the Ricci curvature vanishes, which is what we wanted.

We should note that the notion of balanced discussed in \cite{Scalar}, for which the asymptotic expansion derived here makes sense, does not agree with Definition \ref{ref35}. This is because the volume form defining balancedness also depends on the line bundle metric, and is not fixed to be $\Omega \wedge \bar{\Omega}$ throughout (as a result, it is a more `nonlinear' notion). On a CY the `nonlinear' balanced metric converges to the Ricci-flat one. Thus the asymptotic expansion \ref{ref33} should also hold for our case in an appropriate limit.

\bibliographystyle{amsplain}
\bibliography{biblio}
\end{document}